\documentclass{emulateapj}
\usepackage{graphicx,natbib}
\usepackage{amssymb}
\newcommand{\be}{\begin{equation}}
\newcommand{\ba}{\begin{eqnarray}}
\newcommand{\ee}{\end{equation}}
\newcommand{\ea}{\end{eqnarray}}

\def\lesssim{\mathrel{\hbox{\rlap{\hbox{\lower4pt\hbox{$\sim$}}}\hbox{$<$}}}}
\def\gtrsim{\mathrel{\hbox{\rlap{\hbox{\lower4pt\hbox{$\sim$}}}\hbox{$>$}}}}

\def\simless{\mathbin{\lower 3pt\hbox
   {$\rlap{\raise 5pt\hbox{$\char'074$}}\mathchar''7218$}}}   
\def\simgreat{\mathbin{\lower 3pt\hbox
   {$\rlap{\raise 5pt\hbox{$\char'076$}}\mathchar''7218$}}}   

\begin{document}

\title{The Kinetic Sunyaev-Zel'dovich Effect from Radiative Transfer 
Simulations of Patchy Reionization}
\author{Ilian T. Iliev$^1$, Ue-Li Pen$^1$, J. Richard Bond$^1$, Garrelt
  Mellema$^{2}$, Paul R. Shapiro$^3$}

\altaffiltext{1}{Canadian Institute for Theoretical Astrophysics, University
  of Toronto, 60 St. George Street, Toronto, ON M5S 3H8, Canada;
  iliev@cita.utoronto.ca} \altaffiltext{2}{Stockholm Observatory, AlbaNova
  University Center, Stockholm University, SE-106 91 Stockholm, Sweden}
\altaffiltext{3}{Department of Astronomy, University of Texas, Austin, TX
  78712-1083, U.S.A.}
\submitted{submitted to ApJ}
\label{firstpage}

\begin{abstract}
We present the first calculation of the kinetic Sunyaev-Zel'dovich (kSZ) 
effect due to the inhomogeneous reionization of the universe based on detailed 
large-scale radiative transfer simulations of reionization. The resulting sky
power spectra peak at $\ell=2000-8000$ with maximum values of 
$[\ell(\ell+1)C_\ell/(2\pi)]_{\rm max}\sim1\times10^{-12}$. The peak scale 
is determined by the typical size of the ionized regions and roughly 
corresponds to the ionized bubble sizes observed in our simulations, 
$\sim5-20$~Mpc. The kSZ anisotropy signal from reionization dominates the 
primary CMB signal above $\ell=3000$. This predicted kSZ signal at arcminute 
scales is sufficiently strong to be detectable by upcoming experiments, like 
the Atacama Cosmology Telescope and South Pole Telescope which are expected to 
have $\sim1'$ resolution and $\sim\mu$K sensitivity. The extended and patchy 
nature of the reionization process results in a boost of the peak signal in 
power by approximately one order of magnitude compared to a uniform reionization 
scenario, while roughly tripling the signal compared with that based upon the
assumption of gradual but spatially uniform reionization.
At large scales the patchy kSZ signal depends largely on the ionizing source 
efficiencies and the large-scale velocity fields: sources which produce photons 
more efficiently yield 
correspondingly higher signals. The introduction of sub-grid gas clumping 
 in the radiative transfer simulations produces significantly more power at 
small scales, and more non-Gaussian features, but has little effect at large 
scales. The patchy nature of the reionization process roughly doubles the 
{\em total} observed kSZ signal for $\ell\sim3000-10^4$ compared to non-patchy 
scenarios with the same total electron-scattering optical depth.
\end{abstract}


\section{Introduction}
\label{intro_sect}

The secondary anisotropies of the Cosmic Microwave Background (CMB)
are emerging as one of the most powerful tools in cosmology. The
small-scale anisotropies are probes of cosmological structures and are
thus a valuable tool for studying their formation and properties. The
key effect generating small-scale CMB anisotropies is the
Sunyaev-Zel'dovich (SZ) effect \citep{1969Ap&SS...4..301Z}, produced
by Compton scattering of the CMB photons on moving free electrons.
When this is due to thermal motions the effect is referred to as
thermal SZ effect (tSZ), while when it is due to the electrons moving
with a net bulk peculiar velocity it is called kinetic SZ effect (kSZ)
\citep{1980MNRAS.190..413S}. The former, coming largely from hot gas
in low-redshift galaxy clusters, is the dominant effect, but has a
different spectrum than the CMB primary anisotropies. The tSZ spectrum
has a characteristic zero at $\sim217$~MHz, so its contribution can in
principle be separated. By contrast, kSZ anisotropies have a spectrum 
identical with that of the primary anisotropies. Separation of its effects 
is possible only by virtue of its different spatial structure. We
consider the kSZ fluctuations as composed of two basic contributions,
one coming from inhomogeneous reionization and the other from the
fully-ionized gas after reionization. The calculation of the patchy
reionization contribution to the kSZ signal based on the first
large-scale radiative transfer simulations of reionization is the main
focus of this paper.  This signal provides a signature of the
character of reionization that will complement other approaches like
observations of the redshifted 21-cm line of hydrogen and surveys of
high-z Ly-$\alpha$ emitters
\citep[e.g.][]{1990MNRAS.247..510S,2000ApJ...528..597T,2002ApJ...572L.123I,
2003MNRAS.341...81I,2003ApJ...596....1C, 2003AJ....125.1006R,
2004ApJ...604L..13S,2004ApJ...608L..77G,2004MNRAS.347..187F,
21cmreionpaper,2005astro.ph.11196M,2006ApJ...646..681S,2006NewAR..50...94B,
2006astro.ph..8032F}. In addition to better understanding of the physics 
of reionization, such studies could provide better constraints on the 
fundamental cosmological parameters, in particular on the primordial power 
spectrum of density fluctuations at smaller scales than are currently 
available. 

The kSZ from a fully-ionized medium has been studied previously by
both analytical and numerical means. When linear perturbation theory is 
used, the effect is associated with the quadratic nonlinearities in the 
electron density current and is usually referred to as the Ostriker-Vishniac 
effect \citep{{1986ApJ...306L..51O},{1987ApJ...322..597V}}. Calculating the
full, nonlinear effect is much more difficult to do analytically,
although some models have been proposed
\citep[e.g.][]{1998PhRvD..58d3001J,2000ApJ...529...12H,
2002PhRvL..88u1301M,2004MNRAS.347.1224Z}.  This effect involves
coupling of very large-scale to quite small-scale density fluctuation
modes and thus requires a very large dynamic range in order to be simulated
correctly. Current simulations are in rough agreement, but have not
quite converged yet
\citep[e.g.][]{2001ApJ...549..681S,2002PhRvL..88u1301M,2004MNRAS.347.1224Z}.
All simulations predict significant enhancement of the small-scale
anisotropies compared to the analytical theory, due to the nonlinear
evolution.

Calculations of the kSZ contribution from patchy reionization are even more 
challenging. In addition to the above difficulties, such calculations also 
require detailed modelling of the radiative transfer of ionizing radiation 
to derive the sizes and distributions of H~II regions in space and time (the
reionization geometry) and how these correlate with the velocity and density 
fields. To date this problem has been studied by only a few recent works. Most 
of these estimates were done by semi-analytical models, which used various 
simplified approaches to model the inhomogeneous reionization 
\citep{1998ApJ...508..435G,
2003ApJ...598..756S, 2005ApJ...630..657Z,2005ApJ...630..643M}. While such 
models are useful since they are cheaper to calculate 
than full radiative transfer simulations, and so allow for fast exploration
of the various scenarios, their results must be checked against full 
detailed simulations in order to ascertain their reliability. The only two 
existing numerical studies of this effect, \citet{2001ApJ...551....3G} and 
\citet{2005MNRAS.360.1063S} used fairly small computational boxes 
($4\,\rm h^{-1} Mpc$ and $20\,\rm h^{-1} Mpc$, respectively). As a
consequence, their results significantly underestimate the kSZ signal, as 
we discuss below.

There are several upcoming experiments which will search for kSZ effect, in
particular the Atacama Cosmology Telescope 
(ACT)\footnote{$\url{http://www.hep.upenn.edu/\!\sim\!angelica/act/act.html}$} 
and South Pole Telescope 
(SPT)\footnote{$\url{http://spt.uchicago.edu/science/index.html}$}, both of 
which are expected to be operational by early 2007. These experiments will 
have $\sim1'$ resolution and $\sim\mu$K sensitivities, which should be 
sufficient to detect the kSZ signal.  

In this paper we present the first calculations of the kSZ effect from patchy
reionization based on large-scale radiative transfer simulations. We simulate
reionization using a $(100\,\rm h^{-1}Mpc)^3$ simulation volume, which is
sufficient to capture the relevant large-scale density and velocity
perturbations, an important improvement over previous efforts. We use the 
ray-tracing code C$^2$-Ray \citep{methodpaper} to follow the radiation from 
all ionizing sources in that volume identified with the resolved halos, which 
are of dwarf galaxy size or larger. The halos and underlying density field are 
provided by a very large N-body simulation with the code PMFAST 
\citep{2005NewA...10..393M}. The results of these simulations on the 
reionization character, geometry and observability of the redshifted 21-cm 
line of hydrogen were discussed in detail in 
\citet[][Paper I]{2006MNRAS.369.1625I} and \citet[][Paper II]{21cmreionpaper}.
A preliminary version of our current results was presented in 
\citet{2006astro.ph..7209I}. 
 
\section{Simulations}
\label{simul_sect}

Our simulations follow the evolution of a comoving simulation volume
of $(100\,h^{-1}\rm Mpc)^3$, corresponding to an angular size $\sim1$~deg 
on the sky at the relevant redshift range. Our methodology and simulation 
parameters were described in detail in Papers I and II. 
Here we provide just a brief summary. We start with performing a very 
large pure dark matter simulation of early structure formation, with 
$1624^3\approx4.3$ billion particles and $3248^3$ grid 
cells~\footnote{$3248=N_{nodes}\times(512-2\times24)$, where $N_{nodes}=7$ 
(with 4 processors each), 512 cells is the Fourier transform size and 24 
cells is the buffer zone needed for correct force matching on each side of 
the cube.} using the code 
PMFAST \citep{2005NewA...10..393M}. This allows us to reliably identify 
(with 100 particles or more per halo) all halos with masses
$2.5\times10^9M_\odot$ or larger. We find and save the halo
catalogues, which contain the halo positions, masses and detailed
properties, in up to 100 time slices starting from high redshift
($z\sim30$) to the observed end of reionization at $z\sim6$. We also
save the corresponding density and bulk peculiar velocity fields at
the resolution of the radiative transfer grid. Unfortunately,
radiative transfer simulations at the full grid size of our N-body
computations are impractical on current computer hardware, thus
we solve for the radiative transfer on coarser grids, of sizes 
$203^3=(3248/16)^3$ or $406^3$. These grid resolutions allow us to derive 
reliably the angular sky power spectra for $l\approx 430-90,000$ for $203^3$, 
and up to $\ell\sim180,000$ for $406^3$. Throughout this study we assume a 
flat $\Lambda$CDM cosmology with parameters 
($\Omega_m,\Omega_\Lambda,\Omega_b,h,\sigma_8,n)=(0.27,0.73,0.044,0.7,0.9,1)$
\citep{2003ApJS..148..175S}, where $\Omega_m$, $\Omega_\Lambda$, and
$\Omega_b$ are the total matter, vacuum, and baryonic densities in units of
the critical density, $h$ is the Hubble constant in units of 100
$\rm km\,s^{-1}Mpc^{-1}$, $\sigma_8$ is the standard deviation of linear 
density fluctuations at present on the scale of $8 \rm h^{-1}{\rm Mpc}$, 
and $n$ is the index of the primordial power spectrum. We use the CMBfast 
transfer function \citep{1996ApJ...469..437S}. All calculations are done in 
flat-sky approximation, which is appropriate for our relatively small 
angular sizes. 

All identified halos are assumed to be sources of ionizing radiation and 
each is assigned a photon emissivity proportional to its total mass, $M$, 
according to
\be 
  \dot{N}_\gamma=f_\gamma\frac{M\Omega_b}{\mu m_p t_s\Omega_0},
\ee
where $t_s$ is the source lifetime, $m_p$ is the proton mass, $\mu$ is the 
mean molecular weight and $f_\gamma$ 
is a photon production efficiency which includes the number of photons 
produced per stellar atom, the star formation efficiency (i.e. what fraction
of the baryons are converted into stars) and the escape fraction (i.e. how
many of the produced ionizing photons escape the halos and are available to
ionize the IGM).

The radiative transfer is followed using our fast and accurate ray-tracing
photoionization and non-equilibrium chemistry code C$^2$-Ray. The code has 
been tested in detail for correctness and accuracy against available analytical
solutions and a number of other cosmological radiative transfer codes
\citep{methodpaper,2006MNRAS.tmp..873I}. The radiation is traced from every 
source on the grid to every cell.
 
We have performed four radiative transfer simulations. These share the source 
lists and density fields given by the underlying N-body simulation, but adopt 
different assumptions about the source efficiencies and the sub-grid density 
fluctuations. The runs and notation are the same as in Paper II: runs f2000 
and f250 assume $f_\gamma=2000$ and 250, respectively, and no sub-grid gas 
clumping, while f2000C and f250C adopt the same respective efficiencies, 
$f_\gamma=2000$ and 250, but also add a sub-grid gas clumping, 
$C(z)=\langle n^2\rangle/\langle n\rangle^2$, which evolves with redshift 
according to
\be
C_{\rm subgrid}(z)=27.466 e^{-0.114z+0.001328\,z^2}.
\label{clumpfact_fit}
\ee
The last fit was obtained from another high-resolution PMFAST N-body 
simulation, with box size $(3.5\,\rm h^{-1}~Mpc)^3$ and a computational 
mesh and number of particles of $3248^3$ and $1624^3$, respectively. These 
parameters correspond to a particle mass of $10^3M_\odot$ and minimum 
resolved halo mass of $10^5M_\odot$. This box size was chosen so as to 
resolve the scales most relevant to the gas clumping - on scales smaller than
these the gas fluctuations would be below the Jeans scale, 
while on larger scales the density fluctuations are already present in our 
computational density fields and should not be doubly-counted. The expression
in equation~(\ref{clumpfact_fit}) excludes the matter inside collapsed 
minihalos (halos which are too small to cool atomically, and thus have
inefficient star formation) since these are shielded, unlike the generally 
optically-thin IGM. This self-shielding results in a lower contribution of 
the minihalos to the total number of recombinations than one would infer 
from a simple gas clumping argument
\citep{2004MNRAS.348..753S,2005MNRAS...361..405I,2005ApJ...624..491I}. The 
effect of minihalos
could be included as sub-grid physics as well, see \citet{MH_sim}. This 
results in slower propagation of the ionization fronts and further delay 
of the final overlap. The halos that can cool atomically are assumed here 
to be ionizing sources and their recombinations are thus implicitly included 
in the photon production efficiency $f_\gamma$ through the corresponding 
escape fraction.

\section{kSZ from Patchy Reionization}
\label{kSZ_sect}

The kSZ effect is the CMB temperature anisotropy along a line-of-sight
(LOS) defined by a unit vector ${\bf n}$ induced by Thomson scattering
from flowing electrons: 
\be \frac{\Delta T}{T_{\rm CMB}} =\int
d\eta e^{-\tau_{\rm es}(\eta)}an_e\sigma_T{\bf n} \cdot {\bf v},
\label{ksz_int}
\ee
where $\eta=\int_0^t dt'/a(t')$ is the conformal time, $a$ is the scale
factor, $\sigma_T=6.65\times10^{-25}\,\rm cm^{-2}$ is the Thomson scattering
cross-section, and $\tau_{\rm es}$ is the corresponding optical depth.

We calculate the kSZ anisotropy signal from our simulation data as follows. 
We first calculate the line-of-sight integral in equation~(\ref{ksz_int}) for 
each individual output time-slice of the radiative transfer simulation and for 
all LOS along each of the three box axes. The contribution to the total LOS 
integral from each light-crossing time of the box is then obtained by linear 
interpolation between the nearest results from the simulation output times. 
The individual light-crossing time contributions are then all added together 
to obtain the full LOS integral given in equation~(\ref{ksz_int}). All these 
integrals are done for each LOS through the box along the direction of light 
propagation, which allows us to produce kSZ maps, in addition to the 
statistical signals. In order to avoid artificial amplification of the
fluctuations resulting from repeating the same structures along the 
line of sight, after each light-crossing time we randomly shift the box 
in the directions perpendicular to the LOS and rotate the box, so that 
the LOS cycles the directions along the x, y and z axes of the simulation 
volume.

For comparison we also consider two simplified cases in addition to our 
simulations, the cases of instant and of uniform reionization. We define 
instant reionization as a sharp transition from completely neutral to
fully-ionized IGM at redshift $z_{\rm instant}$. We pick
$z_{\rm instant}=13$, which yields the same integrated electron
scattering optical depth as our simulation f250. We also define a
``uniform reionization'' scenario to be one that has the same
time-dependent reionization history (and hence, also the same
$\tau_{\rm es}$) as simulation f250, but spatially uniform (i.e. not
patchy). We then derive the kSZ temperature fluctuations for these 
two scenarios using the same density and velocity data and same
procedures as for the actual simulations. We consider these simplified models
in order to demonstrate the effects of reionization being extended and
patchy in nature.

\begin{figure*}[!ht]
\includegraphics[width=3.5in]{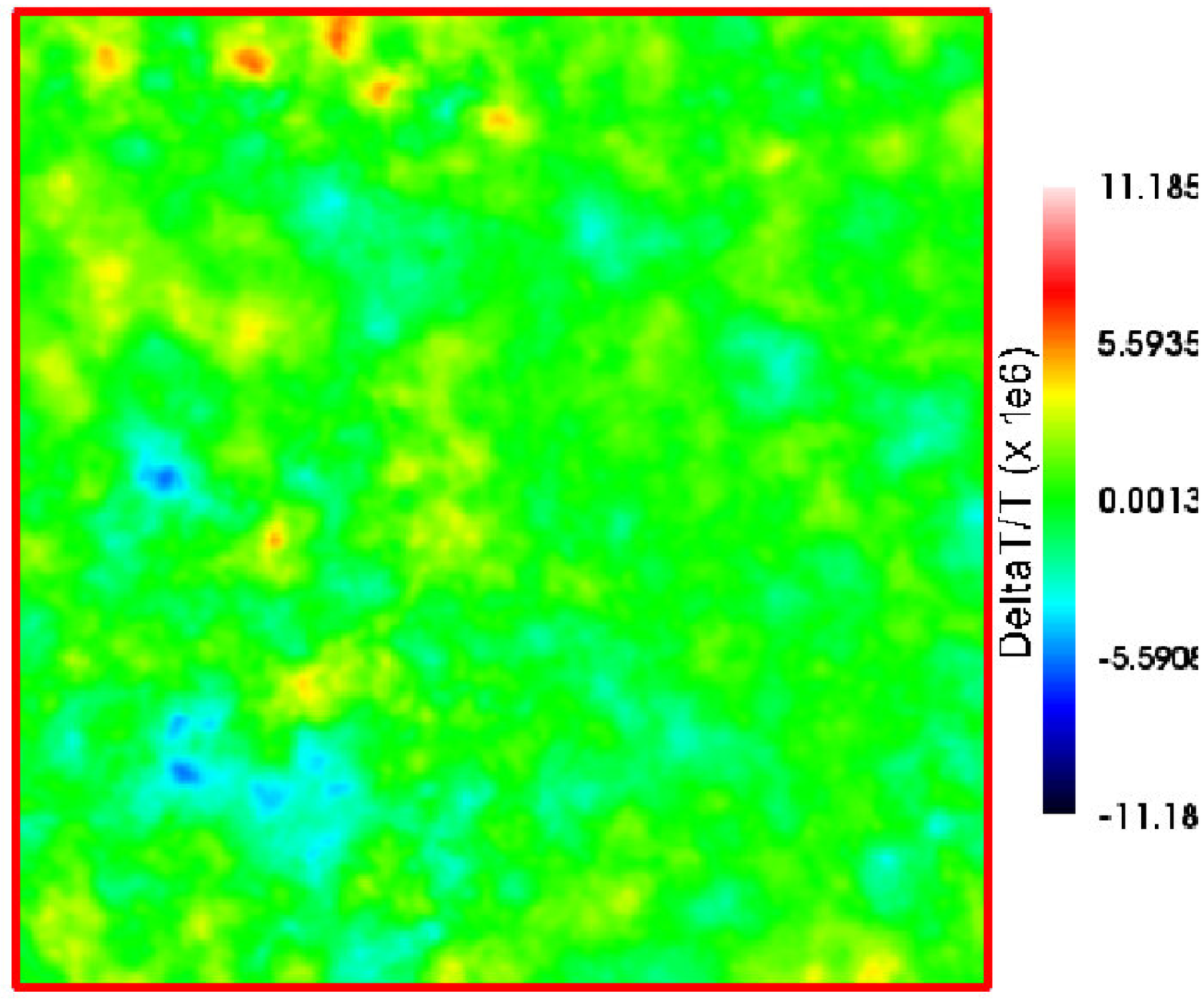}
\includegraphics[width=3.5in]{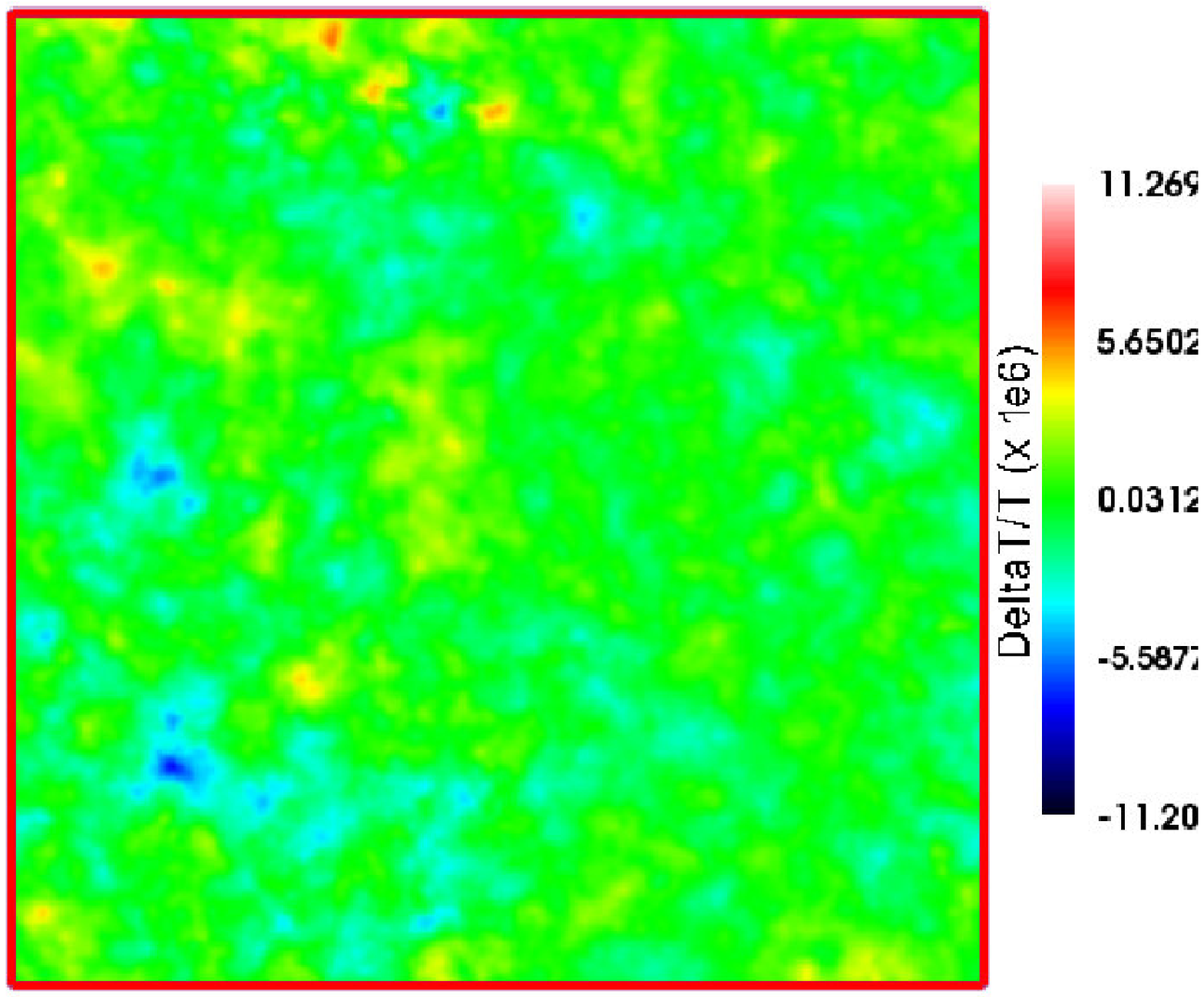}
\includegraphics[width=3.5in]{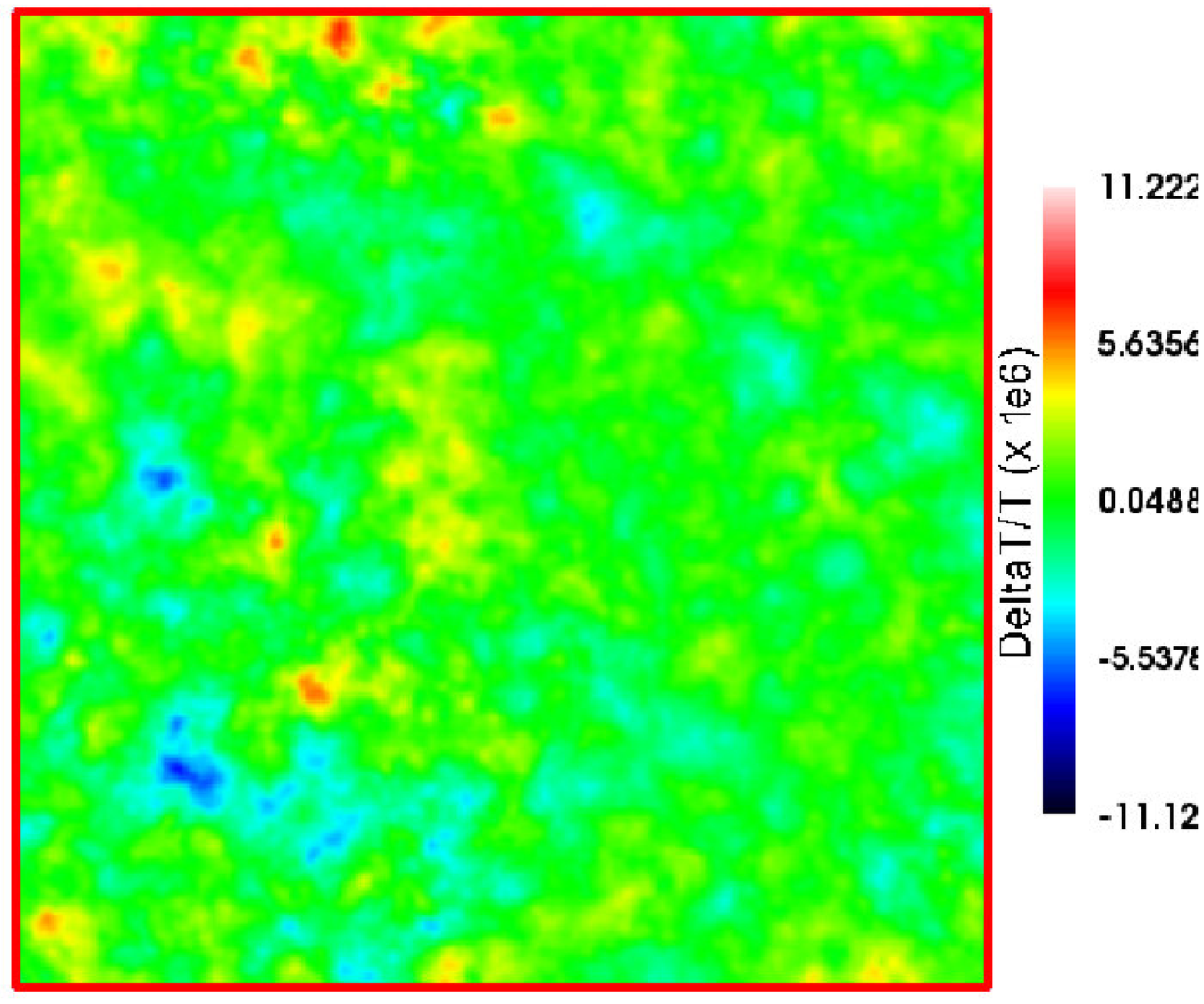}
\includegraphics[width=3.5in]{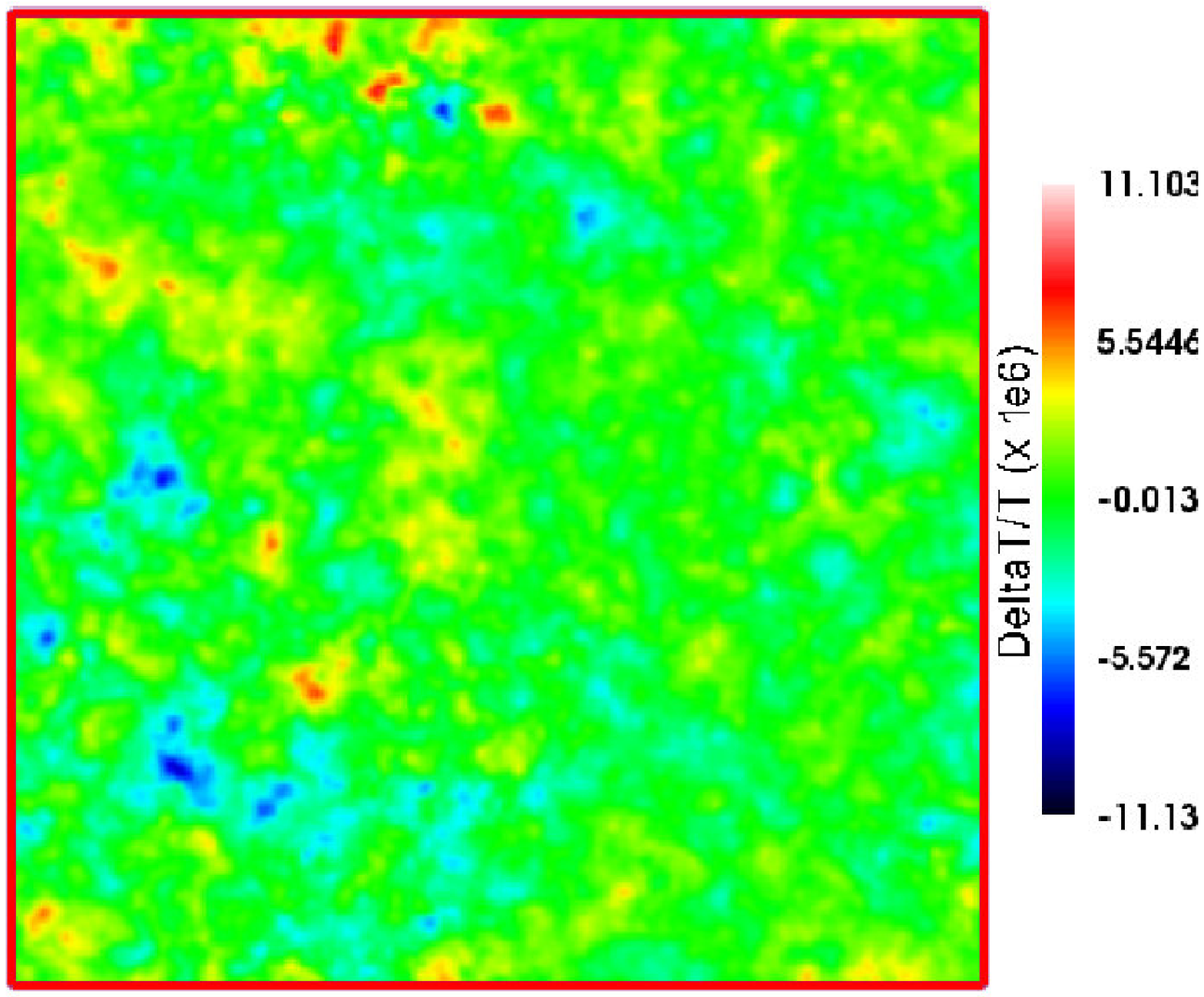}
\includegraphics[width=3.5in]{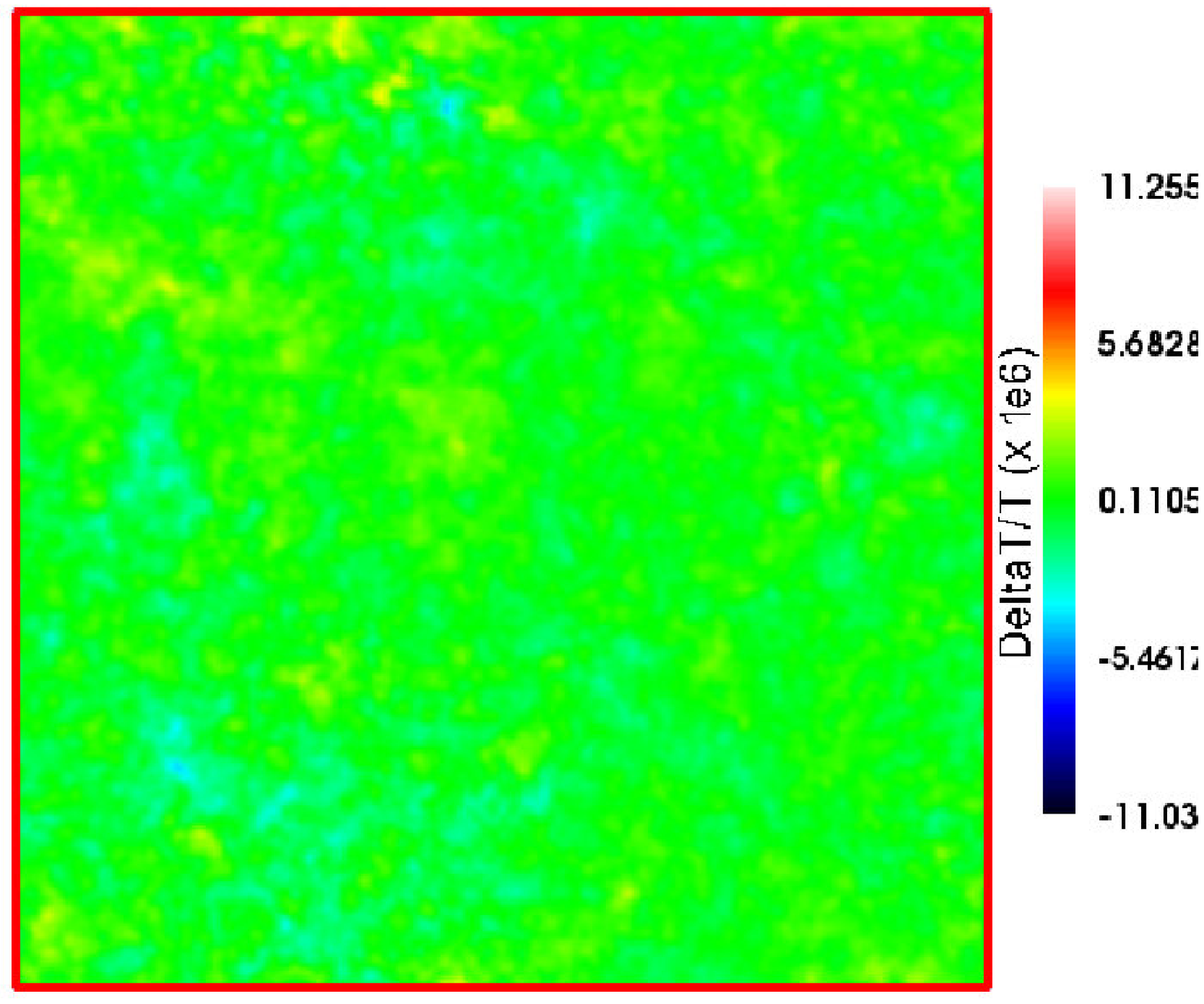}
\includegraphics[width=3.5in]{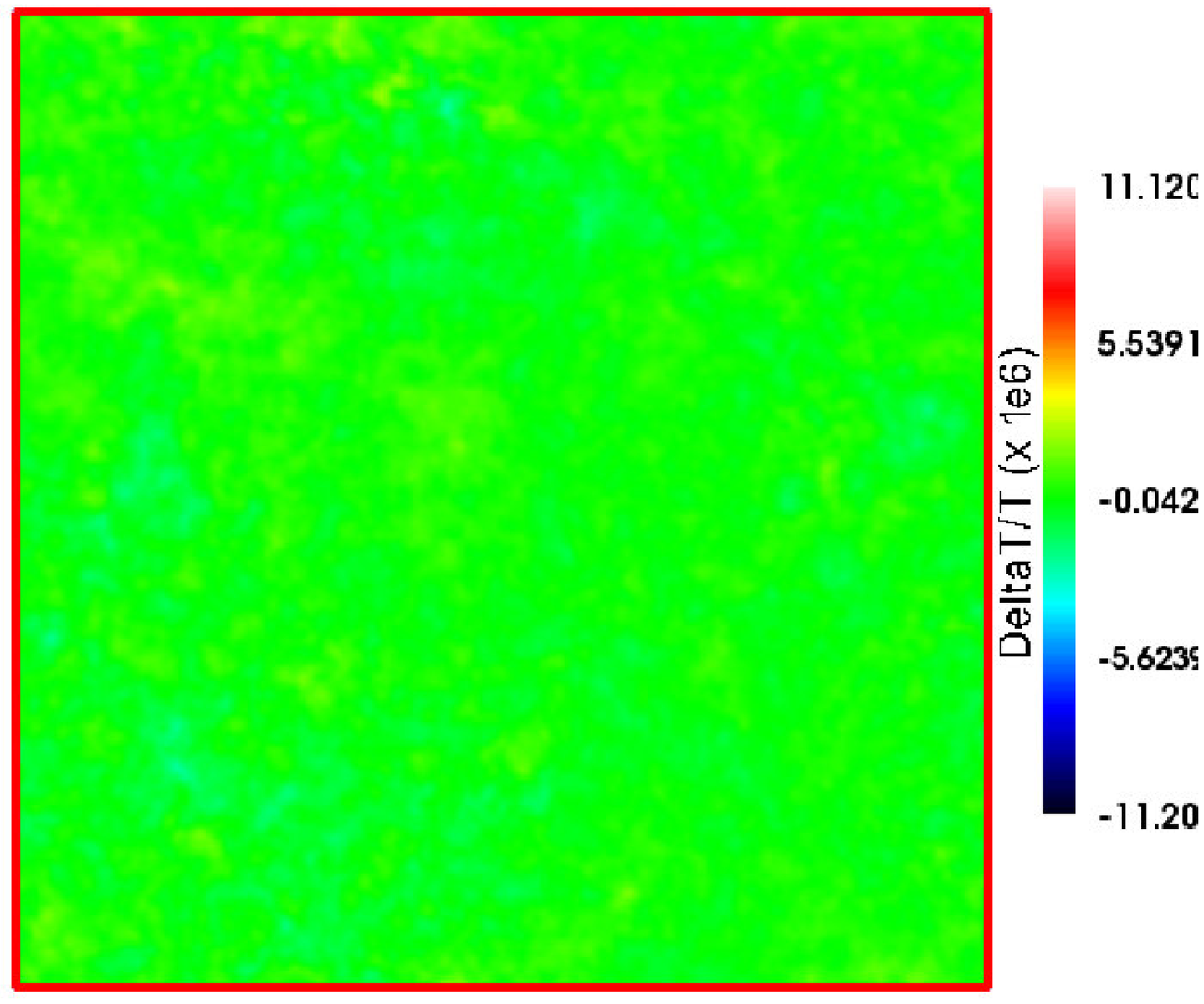}
\caption{\label{maps} kSZ maps from simulations:  f2000 (top left),
f250 (top right), f2000C (middle left), f250C (middle right)
instant reionization at $z_{\rm instant}=13$ which gives the same 
  total electron scattering optical depth as simulation f250 (bottom 
left), and spatially-uniform reionization with the same reionization 
  history and thus same total electron scattering optical depth as 
  simulation f250 (bottom right). (Images produced using the Ifrit 
visualization package of N. Gnedin). 
}
\end{figure*}

\begin{table*}
\caption{Mean and rms values for $\delta T_{\rm kSZ}/T_{\rm CMB}$.}
\label{rms}
\begin{center}
\begin{tabular}{@{}lllllll}
%
simulation & f2000  &                 f2000C    &             f250        &             f250C & uniform & instant\\[2mm]\hline
mean    & $8.86\times10^{-8}$ & $9.88\times10^{-8}$ & $-2.17\times10^{-7}$&$5.57\times10^{-8}$&$-3.29\times10^{-12}$&$-3.37\times10^{-12}$\\[2mm] 
rms     & $1.13\times10^{-6}$ & $1.22\times10^{-6}$ & $1.06\times10^{-6}$ &$1.09\times10^{-6}$&$4.24\times10^{-7}$&$6.67\times10^{-7}$\\[2mm]
\hline\\
\end{tabular}
\end{center}
\end{table*}

\section{Results}

\subsection{kSZ maps}
We show the kSZ maps of temperature fluctuations, $\delta T/T_{\rm
CMB}$, yielded by all of our cases in Figure~\ref{maps}. These have 
a total angular size of
approximately $50'\times50'$, corresponding to our computational box
size. The resolution of the maps is that of the full radiative
transfer grid ($203\times203$ pixels), corresponding to pixel
resolution of $\sim0.25'$. All maps utilize the same color map in
order to facilitate their direct comparison. The maps deriving from
our simulations of inhomogeneous reionization (Figure~\ref{maps}, top 
and middle) all show fairly strong fluctuations, both positive and
negative, of order $\delta T\sim10\, \mu K$ at angular scales of a few
arcminutes ($\rm\sim10~h^{-1}Mpc$) and up to $\delta T\sim20\, \mu K$
at the full map resolution. The fluctuations are at somewhat smaller
scales when the sources are less efficient photon producers (f250 and
f250C), compared to the high-efficiency cases (f2000 and f2000C). The
variations are noticeably enhanced when sub-grid gas clumping is
included in the reionization model (f2000C and f250C) and there are a
number of regions with very strong features. In comparison, the
artificial models of instant and uniform reionization
(Figure~\ref{maps}, bottom) show fluctuations with much lower amplitudes and
with a less well-defined typical scale. Since these simple scenarios
were constructed to produce the same total electron scattering optical
depth as simulation f250 and they use the same density and velocity
fields as the simulations, any observed differences are due to the
patchiness of reionization.  The lower signal in these last two cases
is expected since the kSZ effect in uniformly-ionized gas exactly
cancels in the first order of the linear theory \citep{1984ApJ...282..374K}. 
This cancellation is 
broken when patchiness is present, however, which enhances the anisotropies.

The mean and rms of the temperature fluctuations for all cases are summarized
in Table~\ref{rms}. Based on these, we see that, indeed, the uniform and
instant reionization cases have means very close to zero, as expected,
while the realistic, patchy reionization simulations yield mean temperature 
fluctuations that are low, of order $\sim10^{-7}$, but not zero. The rms 
values are $\sim10^{-6}$ for all patchy reionization cases and lower than 
that by factor of $\sim2$ ($\sim3$) for the instant (uniform) reionization 
cases.

\begin{figure}
\includegraphics[width=3.5in]{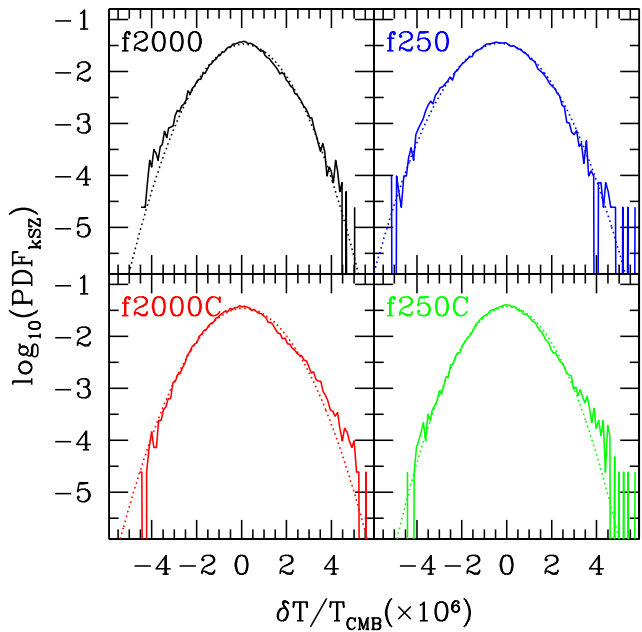}
\caption{\label{PDF_fig} PDF distribution of $\delta T_{\rm kSZ}/T_{\rm CMB}$ 
(solid) vs. Gaussian distribution with the same mean and width (dotted) for 
our four simulations, as labelled.  
}
\end{figure}

\begin{figure}
\includegraphics[width=3.5in]{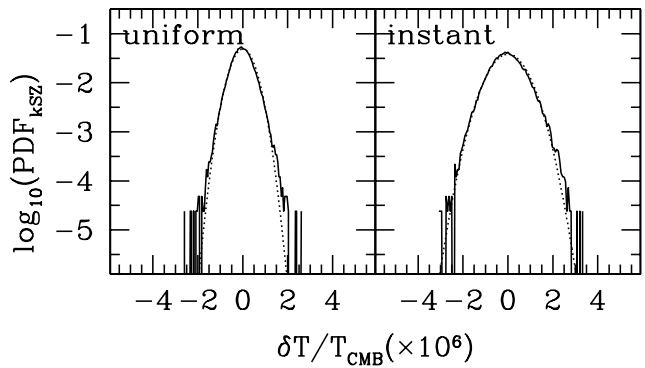}
 \vspace{-1.5in} 
\caption{\label{PDF_special_fig} PDF distribution of 
$\delta T_{\rm kSZ}/T_{\rm CMB}$ (solid) vs. Gaussian distribution with 
the same mean and width (dotted) for the fiducial scenarios of uniform 
(left) and instant reionization (right).  
}
\end{figure}

These observations based on the maps are confirmed by their
corresponding pixel PDF distributions, shown in Figures~\ref{PDF_fig}
and \ref{PDF_special_fig}. On each panel we also plot the Gaussian
distribution with the same mean and standard deviation. All
distributions are surprisingly close to Gaussian around their mean
values given the maps. However noticeable departures from Gaussianity
do occur in the wings of the PDFs. In particular, the PDFs for patchy
reionization with sub-grid clumping (f2000C, and f250C) are
significantly non-Gaussian. For the realization being considered here 
there is an over-abundance of bright regions by up to an order of magnitude
compared to the corresponding Gaussian distributions. The reionization
scenarios without sub-grid clumping show much weaker non-Gaussian
features there. This indicates that the observed PDF of kSZ from
patchy reionization may give us important information on the level of
small-scale gas clumpiness during reionization.  The PDFs derived from
the uniform and instant reionization scenarios are significantly less
wide than the simulated ones (which was also shown by their lower rms
values, as was discussed above), and are also much closer to Gaussian.

\begin{figure}[!ht]
\includegraphics[width=3.5in]{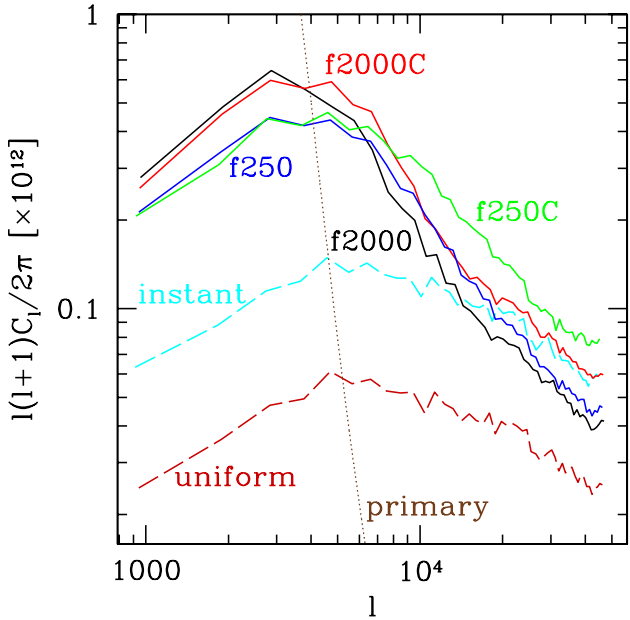}
\caption{\label{ps} Sky power spectra of $\delta T_{\rm kSZ}/T_{\rm CMB}$
  fluctuations resulting from our simulations: f2000 (black), f250
  (blue), f2000C (red) and f250C (green). For comparison, we also show the 
 results from simple models which utilize the same density and velocity fields 
 as the actual simulations, but different assumptions about the gas ionization 
 in space and time, a uniform reionization with the same reionization history
  as simulation f250 (dashed, dark red) and an instant reionization
  model with the same integrated optical depth $\tau_{\rm es}$ as simulation 
 f250 (dashed, cyan). The primary CMB anisotropy signal is also shown 
 (dotted, brown).} 
\end{figure}

\begin{figure*}
\includegraphics[width=2.1in]{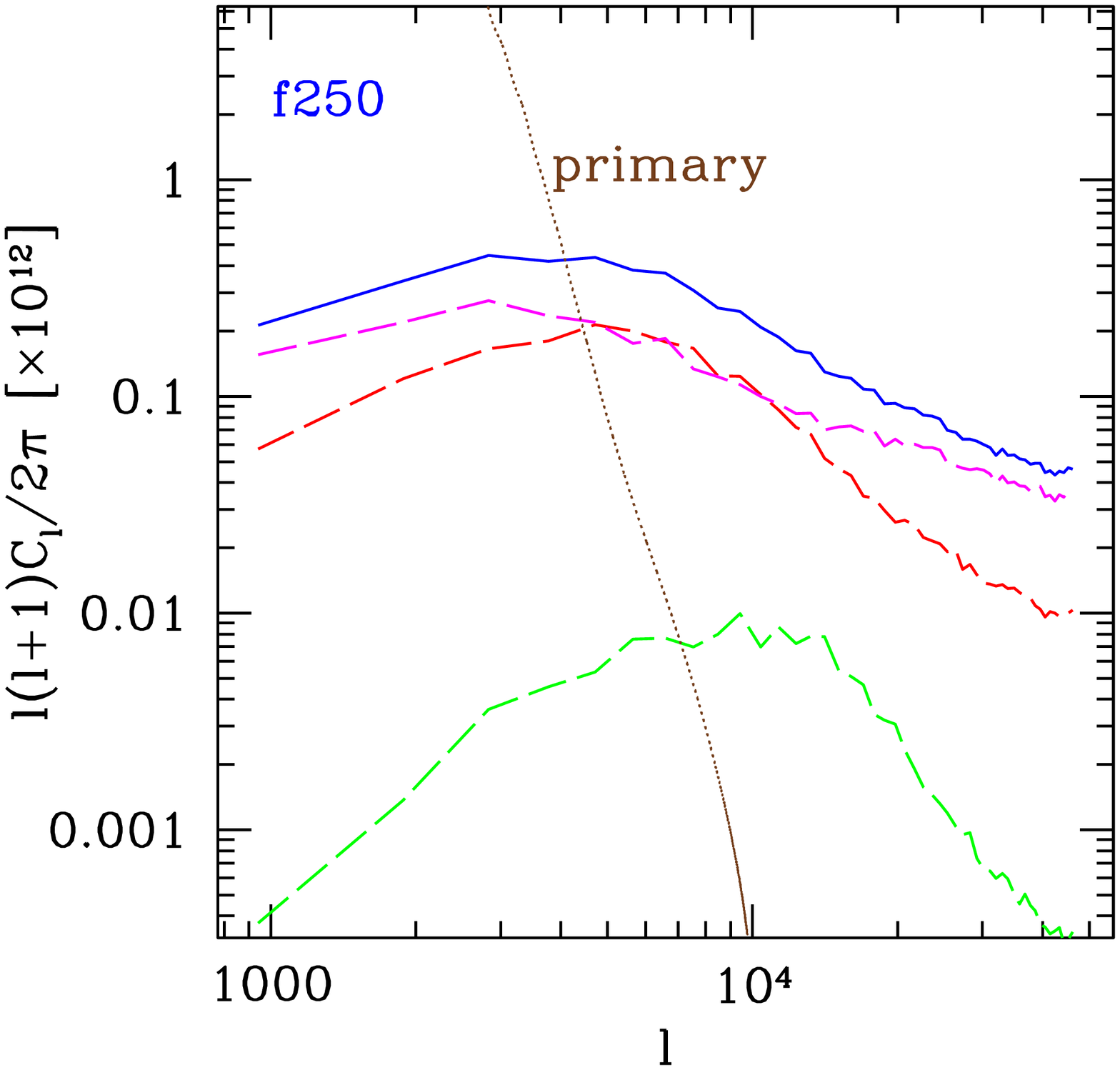}
\includegraphics[width=2.1in]{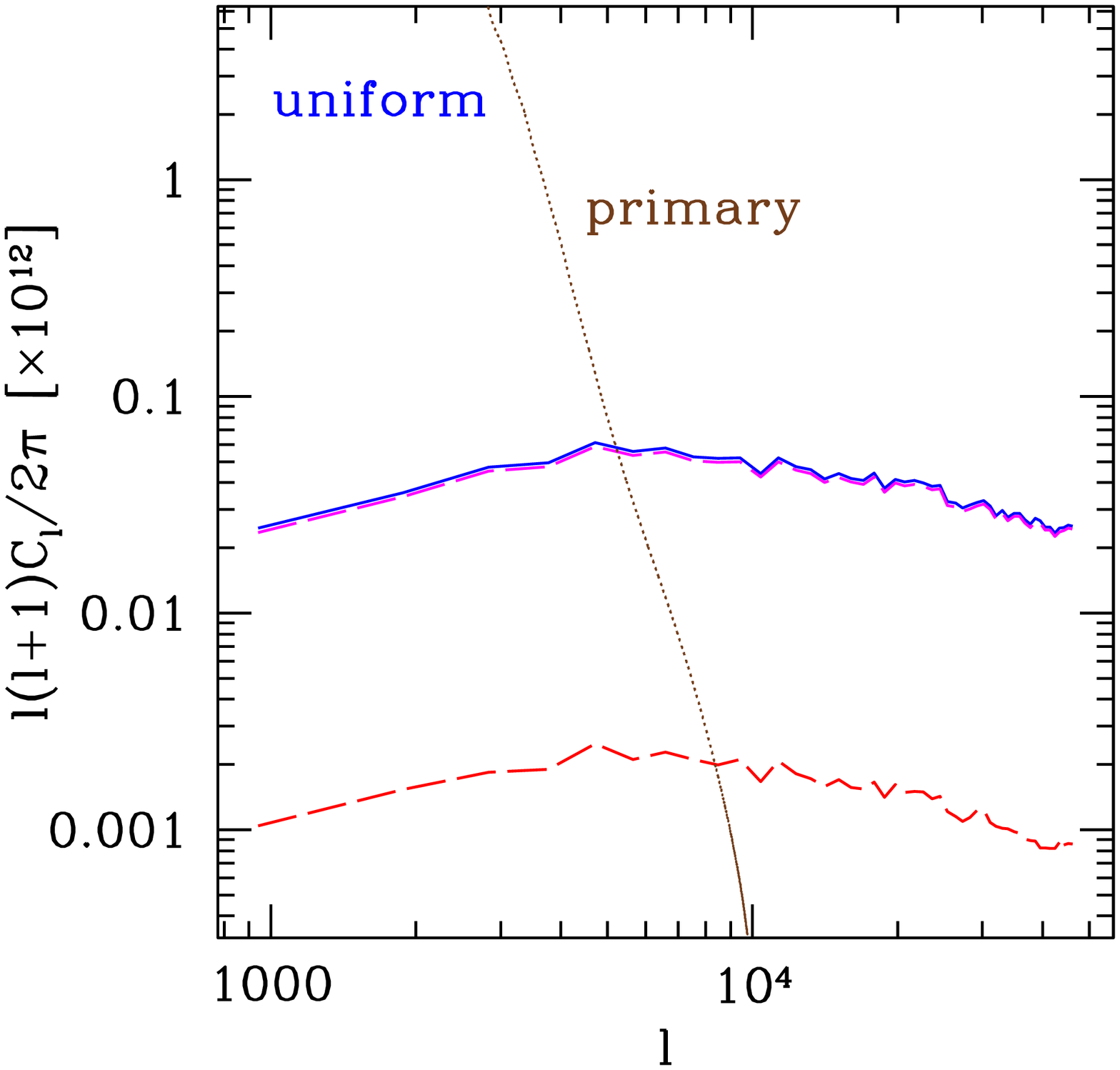}
\includegraphics[width=2.1in]{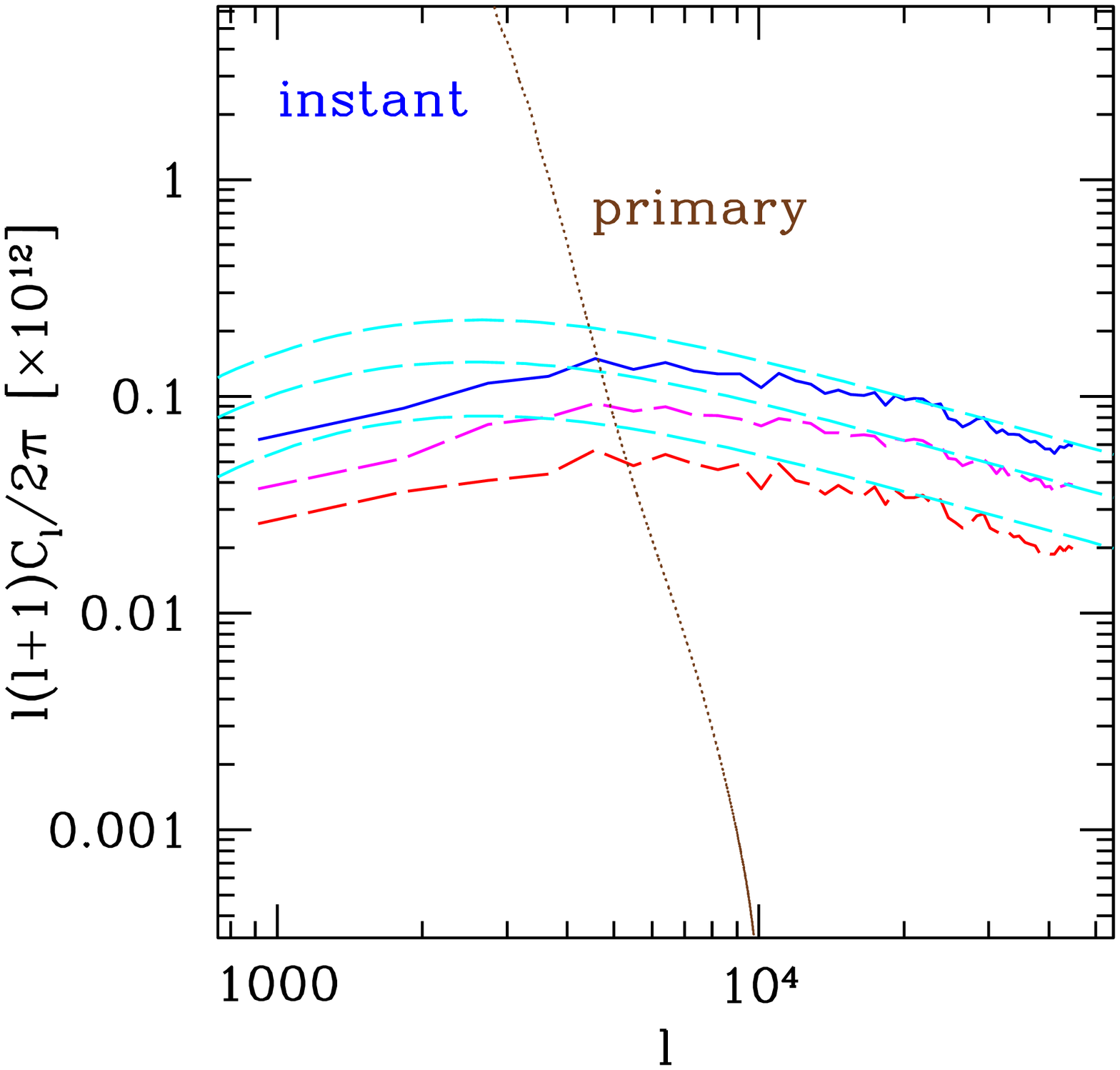}
\caption{\label{fig4} Contributions to the total kSZ signal (blue) from 
different redshifts for case f250 (left), uniform (middle) and instant 
reionization (right). Shown are the signals for every three box light 
crossing times, roughly corresponding to (bottom to top on the left) 
redshifts $z>15$ (green), $15>z>11$ (red), $z<11$ (magenta). In the last 
two cases the $z>15$ contribution is very low (uniform) or zero (instant) 
and thus not show. For comparison, on the last panel we also show the 
corresponding redshift-binned results from the model of 
\citet{2004MNRAS.347.1224Z} (see \S~\ref{lin_th_and_corr_sect}).}
\end{figure*}

\subsection{Sky power spectra}
In Figure~\ref{ps} we show the 2D sky power spectra derived from our 
simulation data. These were obtained by calculating the sky power 
spectra for each box light-crossing time and adding these, which yields
a smoother final result than would calculating the power spectrum 
directly from the final map. The kSZ anisotropy signal from inhomogeneous 
reionization dominates the primary CMB anisotropy above $\ell=3000$. 
The sky power spectra peak strongly at $\ell=3000-5000$, to a maximum of
$[\ell(\ell+1)C_\ell/(2\pi)]_{\rm max}\sim7\times10^{-13}$ when the
ionizing sources are highly-efficient (cases f2000 and f2000C). The
peak values for the simulations with lower-efficiency sources (f250
and f250C) are only slightly lower, at
$[\ell(\ell+1)C_\ell/(2\pi)]_{\rm max}\sim4\times10^{-13}$, and the
peaks are somewhat broader and moved to smaller scales,
$\ell\sim3000-7000$, than for the high-efficiency cases. The
introduction of sub-grid gas clumping in the radiative transfer
simulations produces significantly more power at small scales and
slightly broader peaks when compared to the similar cases without
sub-grid clumping. The scales at which the power spectra peak roughly
correspond to the typical ionized bubble sizes observed in our
simulations, namely $\sim5-20$~Mpc comoving. These typical sizes
depend on the assumed source efficiencies and sub-grid clumping. At
large scales the patchy kSZ signal depends only on the source
efficiencies, being higher for more efficient photon emitters, since
these tend to produce larger ionized regions on average.

In contrast, the uniform reionization scenario (with the mean reionization
history of simulation f250) yields a kSZ signature which is much lower than 
the non-uniform reionization scenarios, indicating a very large boost of the
signal due to the effects of patchiness. The boost is largest, approximately 
one order of
magnitude, at and above the typical scale of the patches ($\ell<7000$), but it
still exists at smaller scales, where it is a factor of two or more. The other
simplified  scenario, of instant reionization with the same integrated optical
depth as f250, produces larger kSZ anisotropy than a uniform reionization
does. However, it is still well below the realistic patchy reionization
signals, by factor of $\sim3$ for $\ell<7000$. At smaller scales the
reionization 
signal from the instant reionization scenario becomes similar to the ones from 
simulated non-uniform reionization. The distribution of the power in these two 
simplified scenarios is much flatter, with less indication of a characteristic 
scale. This implies that the sharp peaks yielded by the inhomogeneous
reionization scenarios are dictated by the size of the ionized patches, since
both the density and the velocity fields are shared among our models. This
point is discussed further in \S~\ref{lin_th_and_corr_sect}.

This behavior can be understood further by considering the contributions from
different redshift intervals to the integrated signal. In Figure~\ref{fig4} we 
show the contributions to the total signal from different redshift intervals  
for simulation f250, as well as the uniform and instant reionization scenarios.
We plot the contribution from every three light-crossing times of the box, 
corresponding roughly to $z>15$, $15>z>11$ and $z<11$. The mass-weighted 
ionized fraction in the highest redshift interval is $x_m<0.01$, and
consequently its contribution to the kSZ effect (case f250, bottom curve) is
low, with a maximum of $\sim10^{-14}$. This high-redshift contribution is
strongly peaked at very small scales ($\ell>10^4$), reflecting the fact that
at that time the typical ionized bubble is still fairly small, less than a few
Mpc in size. The contribution from the middle redshift interval, $11<z<15$,
(corresponding to $0.8>x_m>0.01$) peaks at roughly the same scales as the
total signal. This interval contributes about half of the total signal around 
the peak, but a smaller fraction at scales above and below that. A half or
more of the integrated kSZ signal at all scales is contributed by the lowest 
redshifts ($z<11$, $x_m\gtrsim0.8$). The low-redshift power spectrum is fairly 
flat, with a weak peak at relatively large scales, $\ell\sim2000$. This
late-time contribution resembles the results for the two homogeneous
scenarios, but has higher amplitude and a different shape. This
indicates that even at late times, when most of the IGM is already ionized, 
the patchiness still plays an important role in assembling the kSZ signal.

In the homogeneous (uniform and instant) reionization cases the shape of the
power spectra is essentially the same for all redshift intervals, quite flat, 
with a broad peak at $\ell\sim5000$. The signal from the
uniform reionization scenario is completely dominated by the lowest redshifts,
with less than a few per cent coming from the middle redshift interval, and
essentially no contribution from high redshifts. Since the velocity and
density fields are shared, the instant reionization scenario contributions
differ from the corresponding ones from the uniform scenario only by being 
weighted by the ionized
fraction in the uniform case. The instant reionization scenario assumes full
ionization after $z_{\rm instant}=13$ and fully-neutral gas before that. Thus,
as a consequence to the kSZ signal being stronger at later times, the instant
reionization integrated signal is higher as well. For comparison, we also show 
the corresponding redshift-binned results from the model of 
\citet{2004MNRAS.347.1224Z} (see \S~\ref{lin_th_and_corr_sect}) against our 
results from the instant reionization scenario. The two results agree well
at small scales, in the potentially-observable range. 

\subsection{Linear theory and large-scale velocity field corrections}
\label{lin_th_and_corr_sect}

\begin{figure}[!ht]
  \includegraphics[width=3.5in]{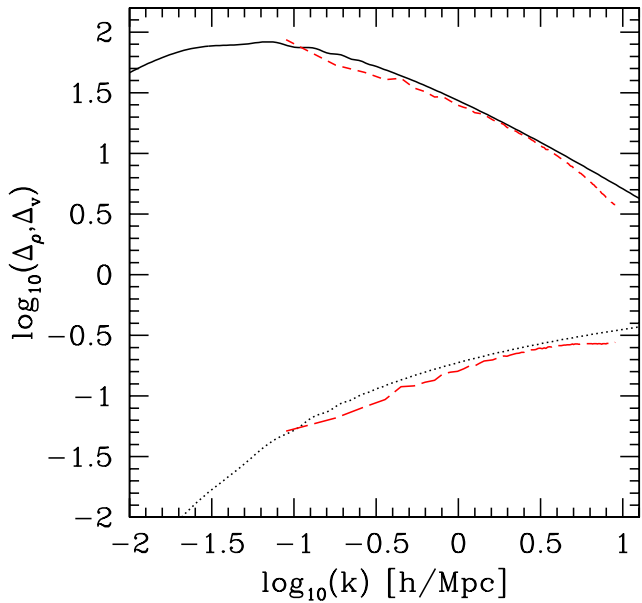}
\caption{\label{ps_lin_sim} Linear-theory power spectra at redshift $z=13.6$ 
of the density field (dotted, black) and velocity field (in $\rm km\,s^{-1}$; 
solid, black) compared to the simulation results for the density (long-dashed, 
red) and velocity (short-dashed, red).  }
\end{figure}

In linear theory the velocity and density perturbations at redshift $z$ are
related through the continuity equation: 
\be 
\tilde{v}({\bf k},t)
=\frac{ia(t)}{k^2}\frac{\dot{D}(t)}{D(t)}{\bf k}\tilde{\delta}({\bf k},t),
\label{lin_vel}
\ee 
where $a=(1+z)^{-1}$ is the scale factor, and $D(t)$ is the growth factor of
linear perturbations, which for a flat $\Lambda$CDM background cosmology is 
given by 
\be
D(z)=\frac{5\Omega_0E(z)}{2}\int_z^\infty\frac{1+z'}{[E(z')]^3}dz',
\label{growth}
\ee 
with $E(z)=H(z)/H_0=[\Omega_0(1+z)^3+\Omega_\Lambda]^{1/2}$, where $H(z)$
and $H_0$ are the values of the Hubble constant at redshift $z$ and at
present, respectively. In terms of the power spectra of the density,
$P_{\delta\delta}$, and the velocity, $P_{vv}$, equation~(\ref{lin_vel})
can be re-written as 
\be
P_{vv}(k,t)
 =\frac{a^2(t)}{k^2}\left[\frac{\dot{D}(t)}{D(t)}\right]^2
      P_{\delta\delta}(k,t),
\ee 
or 
\be
\Delta_{v}^2(k,t)=\frac{a^2(t)}{k^2}\left[\frac{\dot{D}(t)}{D(t)}\right]^2
   \Delta_{\delta}^2(k,t),
\label{lindeltas}
\ee 
where we defined 
\be
\Delta^2_\delta\equiv\frac{k^3}{2\pi^2}P_{\delta\delta}
\label{delta_delta}
\ee 
and
\be
\Delta^2_v\equiv\frac{k^3}{2\pi^2}P_{vv},
\label{delta_v}
\ee 
which would be our notation for the power spectra hereafter.
The quantities $\Delta^2_\delta$ and $\Delta^2_v$ are the power per 
logarithmic interval of the wave-number $k$ of the density and velocity 
fields, respectively.

Using equation~(\ref{growth}) it is straightforward to show that 
\ba
\frac{\dot{D}}{D}&=&\frac{\ddot{a}}{\dot{a}}-\frac{\dot{a}}{a}
+\frac{5\Omega_0}{2}\frac{\dot{a}}{a}\frac{(1+z)^2}{[E(z)]^2D(z)}\nonumber\\
&=&-\frac{3H_0\Omega_0(1+z)^3}{E(z)}\left[1-\frac{5}{3(1+z)D(z)}\right].
\label{dotgrowth}
\ea 
Substituting equation~(\ref{dotgrowth}) into equation~(\ref{lindeltas}) we
finally find 
\ba
&&\Delta^2_v(k,z)=\nonumber\\
&&\,\,\frac{\Delta^2_\delta(k,z)}{k^2}\frac{9H_0^2\Omega_0^2(1+z)^4}{4E^2(z)}
\left[1-\frac{5}{3(1+z)D(z)}\right]^2.
\label{deltavlin}
\ea 
In Figure~\ref{ps_lin_sim} we show the CMBfast density power spectrum at
redshift $z=13.6$, and the corresponding linear-theory velocity power 
spectrum given by equation~(\ref{deltavlin}), along with the density and
velocity power spectra
obtained from our N-body simulation data (re-gridded to our radiative transfer
grid of $203^3$). Comparison of the two sets of data shows a good agreement
between them and indicates that at these scales and redshifts our density and
velocity fields are clearly still in the linear regime. Note, however, that
the bulk velocities derived from the simulation data do include the non-linear
effects at smaller scales, up to the full resolution of the underlying N-body 
simulation, at $3248^3$ cells, or $\sim30\,h^{-1}$~kpc comoving per cell,
averaged at the radiative transfer grid resolution. The velocity power 
spectra were derived from the simulated data by first calculating the power 
spectrum of each of the three velocity components and then averaging these,
which minimizes the variance at large scales.

\begin{figure}
  \includegraphics[width=3.5in]{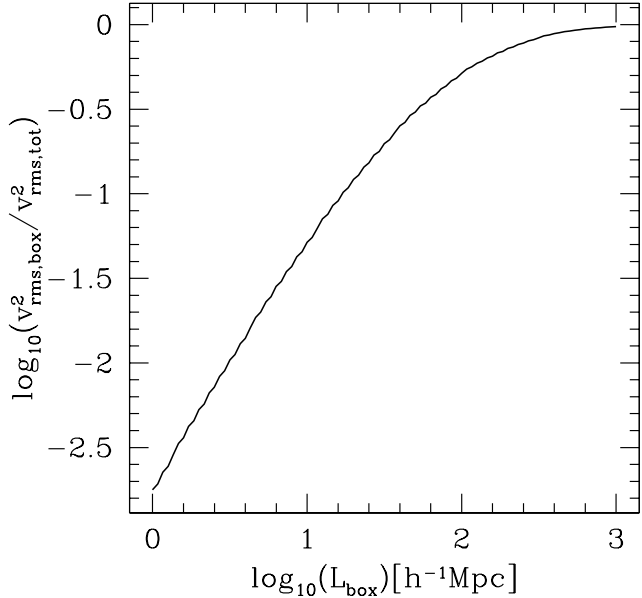}
\caption{\label{vrms_missing_power} Fraction of total linear-theory velocity 
power $v^2_{\rm rms,box}/v^2_{\rm rms,tot}$ present in a simulation box of 
size $L_{\rm box}$ vs.  $L_{\rm box}$.  }
\end{figure}

The velocity power spectrum has a broad peak at fairly large scales,
$k\sim0.01-0.1\,\rm h\,Mpc^{-1}$, corresponding to scales $\sim\,60-600\rm\,
Mpc\,h^{-1}$. Thus, our simulation volume is sufficiently large to reach
this velocity peak, but is still missing some velocity power from the largest 
scales. We estimate the missing power as follows. The rms of the
velocity field is given by 
\be 
v^2_{\rm rms}=\int \Delta^2_v(k)d\ln(k).
\ee 
Thus, integrating over the full linear power spectrum yields the total
power in the velocity field, $v^2_{\rm rms,tot}$ in the linear theory. 
A finite simulation box of size $L_{\rm box}$ would not include the modes 
with wave-numbers below $k_{\rm min}=2\pi/L_{\rm box}$. Integrating 
$\Delta^2_v$ over the wave-numbers
$k>k_{\rm min}$, we obtain the total velocity power for a given box size. The
results are shown in Figure~\ref{vrms_missing_power}, where we plot the
fraction of total linear-theory velocity power $v^2_{\rm rms,box}/v^2_{\rm
  rms,tot}$ present in a simulation box  vs. the box size $L_{\rm box}$.
A simulation of volume $(100\,\rm h^{-1}Mpc)^3$ would thus be missing about
$\sim50\%$ of the total $v^2_{\rm rms}$ velocity power as given by the linear 
theory. Simulations with smaller volumes would miss much larger fraction of 
the velocity power, $\sim70\%$ for $50\,\rm h^{-1}Mpc$ box, $\sim90\%$ for
$20\,\rm h^{-1}Mpc$ box, and over $99\%$ for $4\,\rm h^{-1}Mpc$ box.

The large-scale bulk motions not included in our simulation occur on scales 
of $\sim100$~Mpc or larger, well above the characteristic size of the ionized 
patches ($\sim10$~Mpc or less) 
\citep[see also Figure~\ref{power_spectra}]{2006MNRAS.369.1625I}, thus 
on such scales the reionization patchiness averages out and the ionization 
fraction approaches the mean for the universe. We can
approximately account for the missing large-scale velocity power as follows.
At each light-crossing time we can assume that the whole simulation volume is
moving with some random velocity, $\bar{v}_{\rm box}$. In order to account for
the missing velocity power, the value of the component of this random velocity 
at redshift $z$ along the line-of-sight is given by 
\be
\bar{v}_{\rm box}(z)=
v_{\rm rms, missing}
       (-2\ln q)^{1/2}\cos(2\pi\theta), 
\label{gauss_random}
\ee
where $v_{\rm rms, missing}\equiv[v_{\rm rms,tot}(z)^2-v_{\rm
  rms,box}(z)]^{1/2}$,  and $q$ and $\theta$ are uniformly-distributed random
variables between 0 and 1. The form of equation~(\ref{gauss_random}) ensures 
that
$\bar{v}_{\rm box}(z)$ is a Gaussian-distributed random variable with a zero
mean and rms of $v_{\rm rms, missing}$ \citep{Box-Muller}. In terms of the 
contribution from this redshift to the temperature anisotropy this yields
\be
\left(\frac{\Delta T}{T_{\rm CMB}}\right)_{\rm tot}(z)
   =\left(\frac{\Delta T}{T_{\rm CMB}}\right)_{\rm box}(z)
   +\tau_{\rm es}(z)\frac{\bar{v}_{\rm box}(z)}{c},
\ee
where $\tau_{\rm es}(z)$ is the corresponding contribution to the
total electron-scattering optical depth.
\begin{figure}[!ht]
  \includegraphics[width=3.5in]{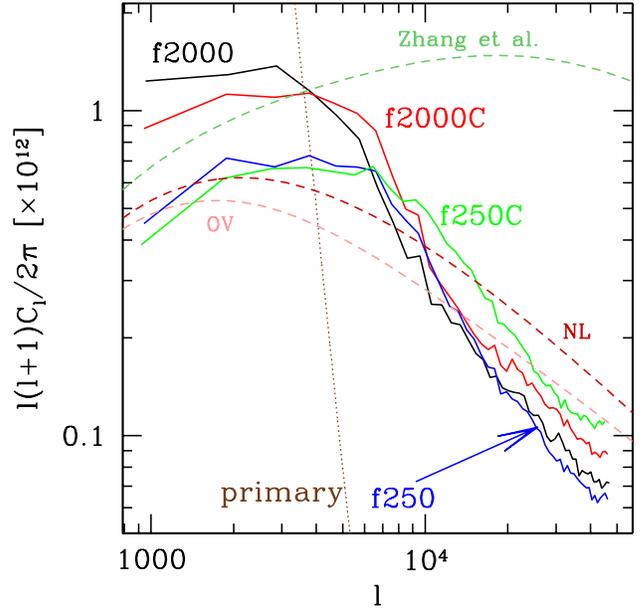}
\caption{\label{ps_w_ran_vel} Sky power spectra of $\delta T_{\rm
  kSZ}/T_{\rm CMB}$ fluctuations from the epoch of reionization based on our 
  simulations, f2000 (black), f250 (blue), f2000C (red) and f250C (green),
  corrected for the missing velocity field power compared to the 
  after-reionization kSZ signals (assuming overlap at $z_{\rm ov}=8$):
  linear Ostriker-Vishniac effect, labelled 'OV' (long-dashed, pink),
  the same, but using the nonlinear power spectrum of density
  fluctuations, labelled 'NL' (long-dashed, dark red) and a
  fully-nonlinear model matched to high-resolution hydrodynamic
  simulations of \citet{2004MNRAS.347.1224Z}, labelled `Zhang et al'
  (short-dashed, dark green).  The primary CMB anisotropy signal is
  also shown (brown, dotted).}
\end{figure}
Note that for uniform reionization the effects from large-scale velocity fields
would exactly cancel, since the flow is potential \citep{1984ApJ...282..374K}. 
However, the patchiness breaks this cancellation and the large-scale velocities
increase the signal by increasing the ionized bubble velocities along the LOS. 
The ionized regions are at much smaller scale than these large-scale 
velocities and uncorrelated with them, thus avoiding the usual cancellation.

\begin{figure*}
  \includegraphics[width=6.5in]{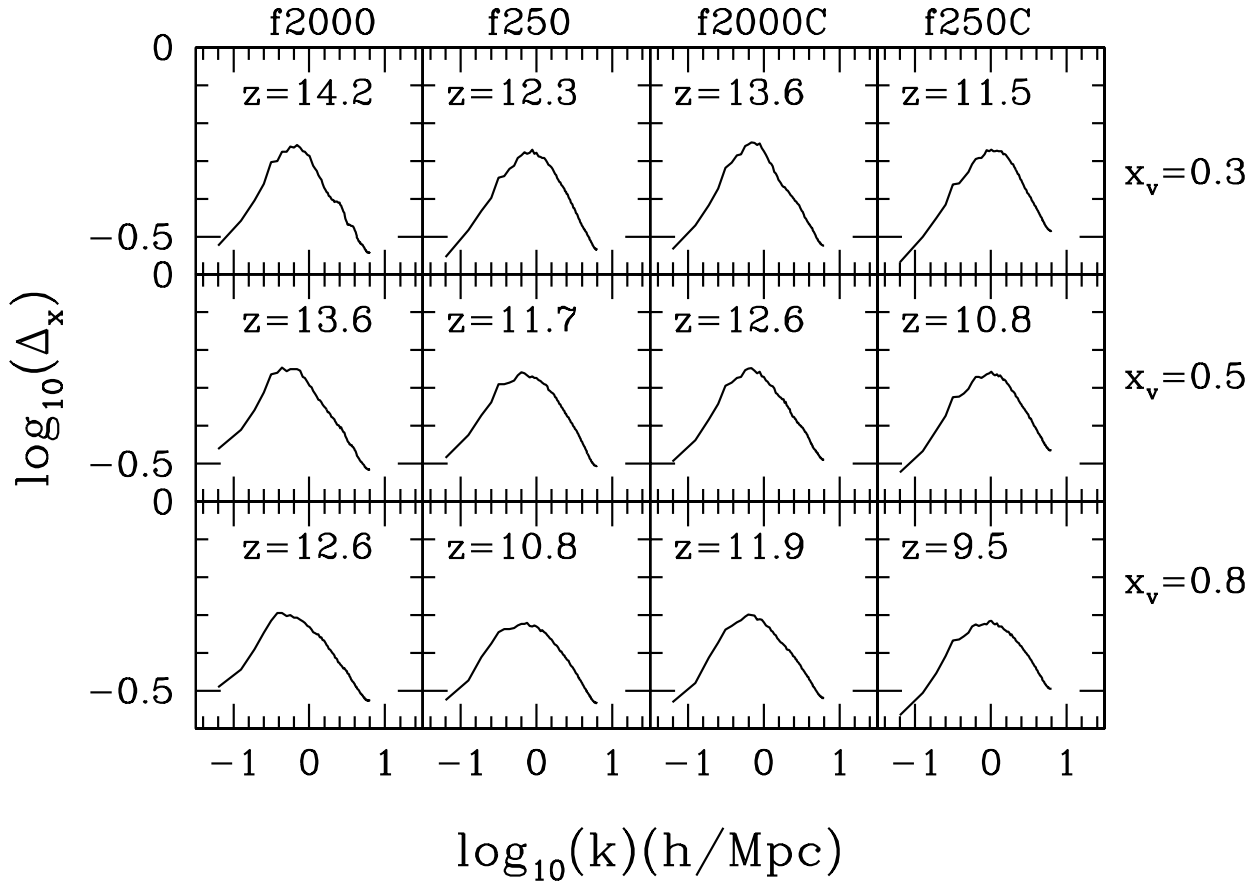}
\vspace{-3.5cm}
\caption{\label{power_spectra} Power spectra, $\Delta_x$ of the 
ionized/neutral fraction at volume ionized fractions $x_v=0.3$ (top), 
$x_v=0.5$ (middle), and $x_v=0.8$ (bottom) for all of our simulations, 
as indicated on top of each column. The corresponding redshifts are 
indicated on each panel.}
\end{figure*}
The power spectra with large-scale velocity corrections included are 
shown in Figure~\ref{ps_w_ran_vel}. For comparison, we also 
show three representative calculations of the post-reionization 
contribution from fully-ionized gas after reionization: the
quadratic-order Ostriker-Vishniac effect \citep{1987ApJ...322..597V}
expressed in terms of a product of the linear power spectra; the same
expression, but with the nonlinear density power spectrum substituted
for one of the linear ones, which partially accounts for the nonlinear 
effects; and the recent detailed nonlinear model of
\citet{2004MNRAS.347.1224Z}. The last three models are rescaled to our
adopted cosmology using $\ell(\ell+1)C_\ell\propto \sigma_8^5$
\citep{2004MNRAS.347.1224Z}, and assume reionization overlap at
$z_{\rm ov}=8$, to allow for direct comparison. Confronting first the
three post-reionization models, we see that they agree fairly well at
large, linear scales, but strongly diverge, by up to one order of
magnitude at small scales, where the nonlinearities become very
important. Including the corrections due to the nonlinear density
power spectrum yields modestly higher kSZ signal than the linear OV
calculation, and similar power spectrum shape overall. Both peak at
roughly the same scale, $\ell\sim2000$. In contrast, the
\citet{2004MNRAS.347.1224Z} result finds much larger signal,
especially at small scales, which peaks at $\ell\sim20,000$.  Compared
to our patchy reionization kSZ predictions, all uniform-ionization
power spectra are much less sharply peaked, i.e. they lack a
well-established characteristic scale. In terms of their peak values,
the OV and OV with nonlinear corrections models yield values lower than
any of our patchy reionization results, while \citet{2004MNRAS.347.1224Z}
find a similar signal around the patchy power spectra peaks. At small 
scales ($\ell>10^4$) the patchy reionization kSZ signals are similar to or 
lower than the OV result and are much lower than the expected full
nonlinear post-reionization contribution. Thus, the best range to aim for 
when attempting to detect the patchy reionization signal is $\ell=3000-10^4$. 

\subsection{Characteristic scales and LOS spectra}

\begin{figure*}
  \includegraphics[width=3.5in]{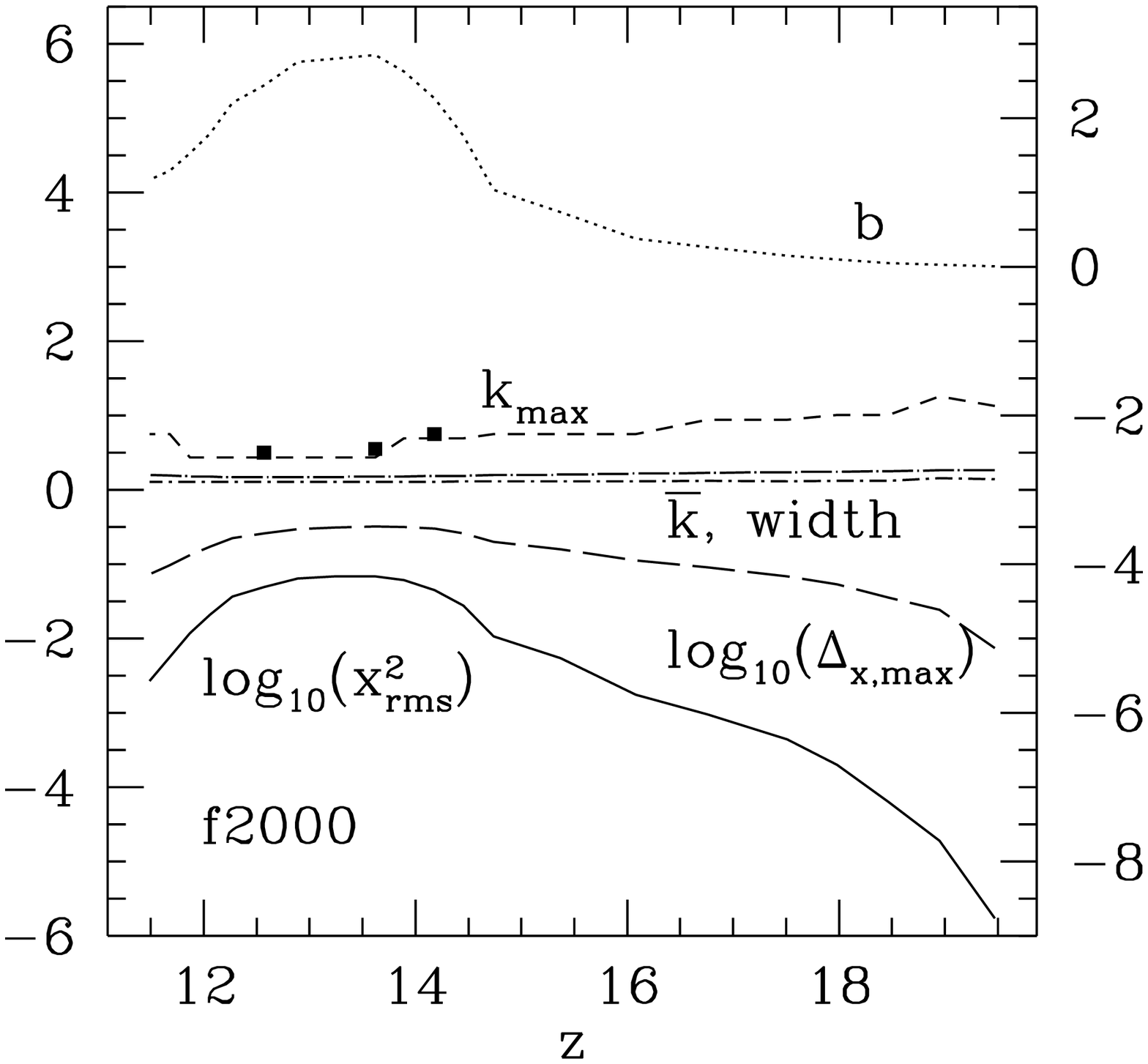}
  \includegraphics[width=3.5in]{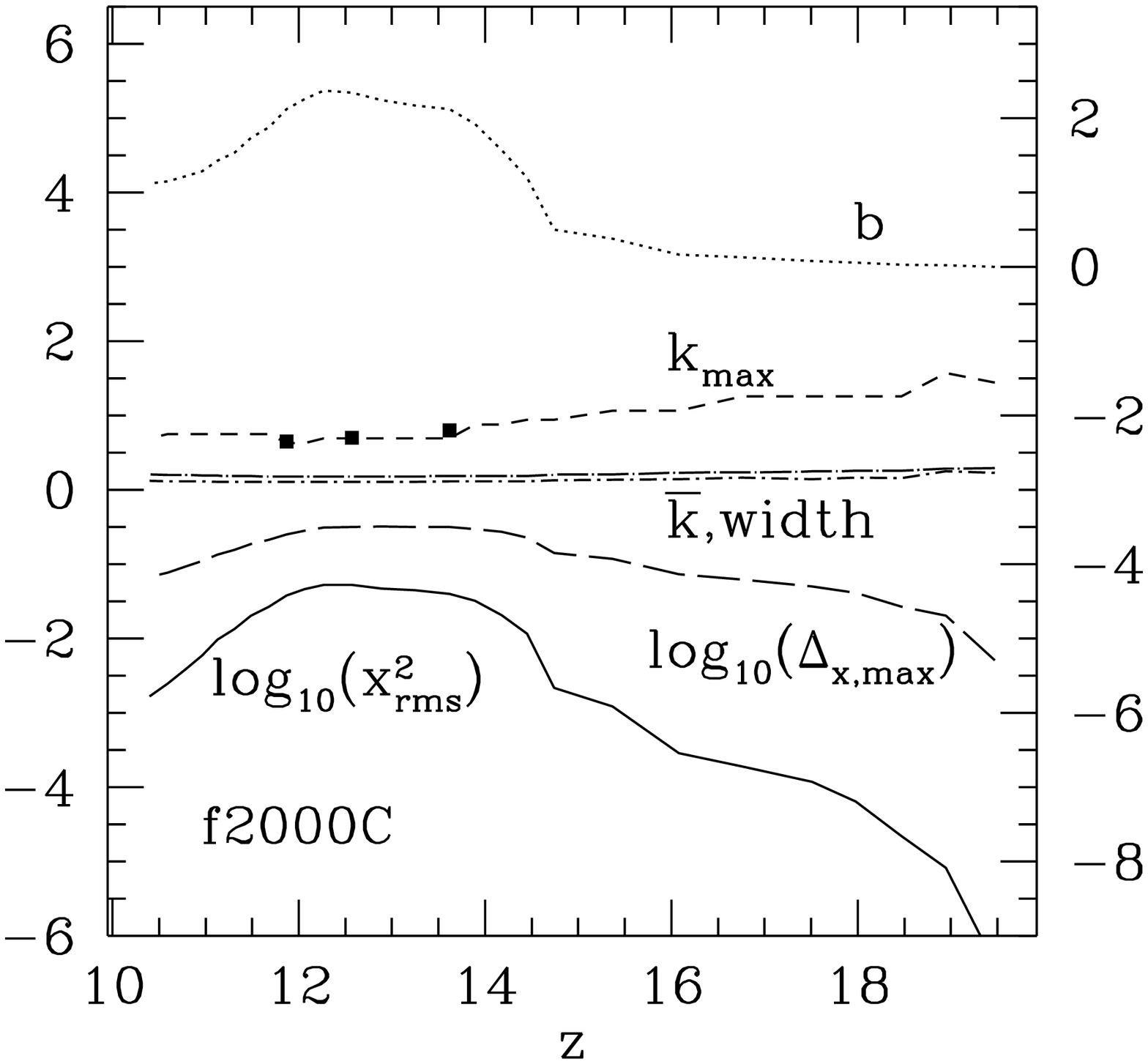}
  \includegraphics[width=3.5in]{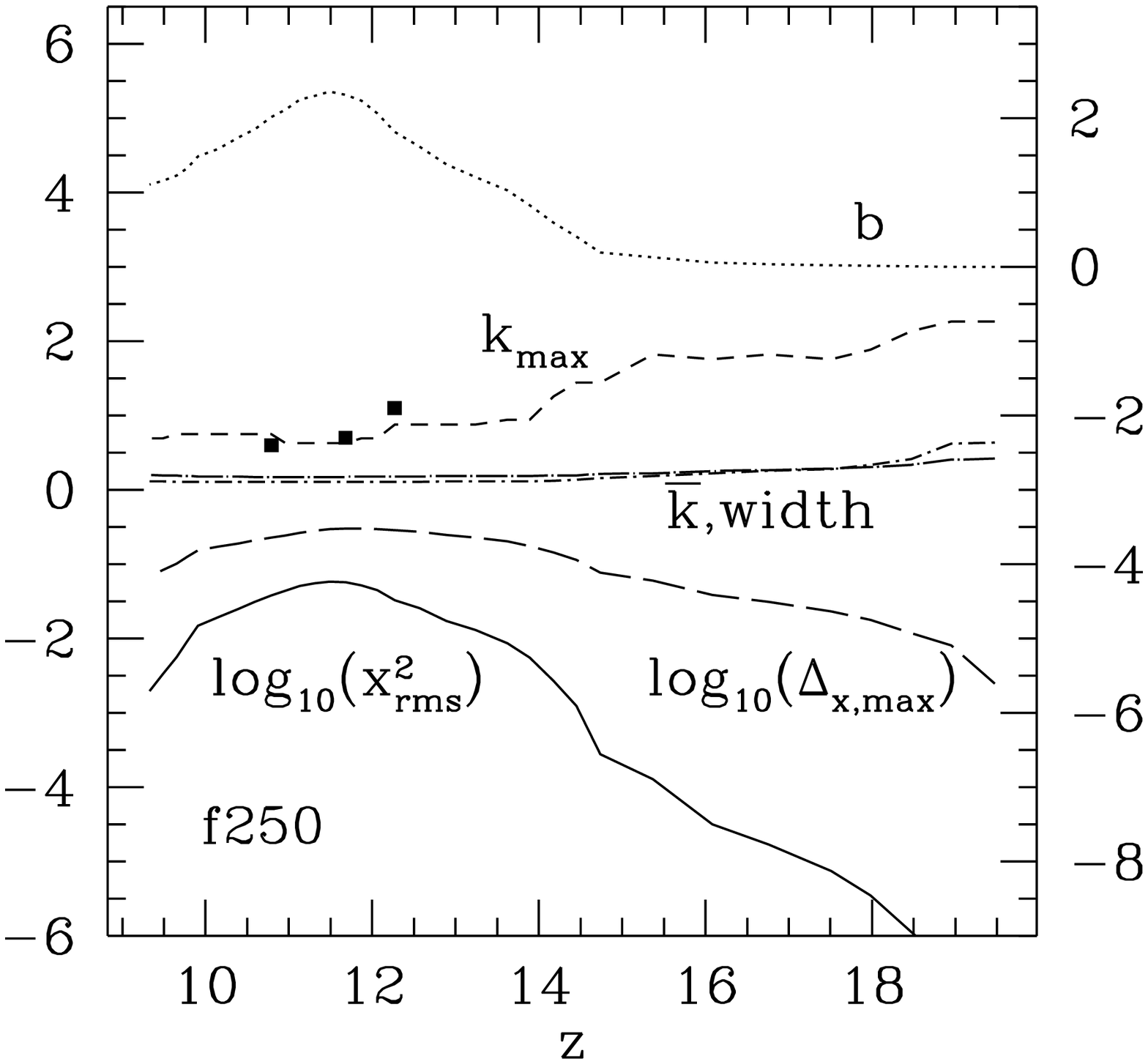}
  \includegraphics[width=3.5in]{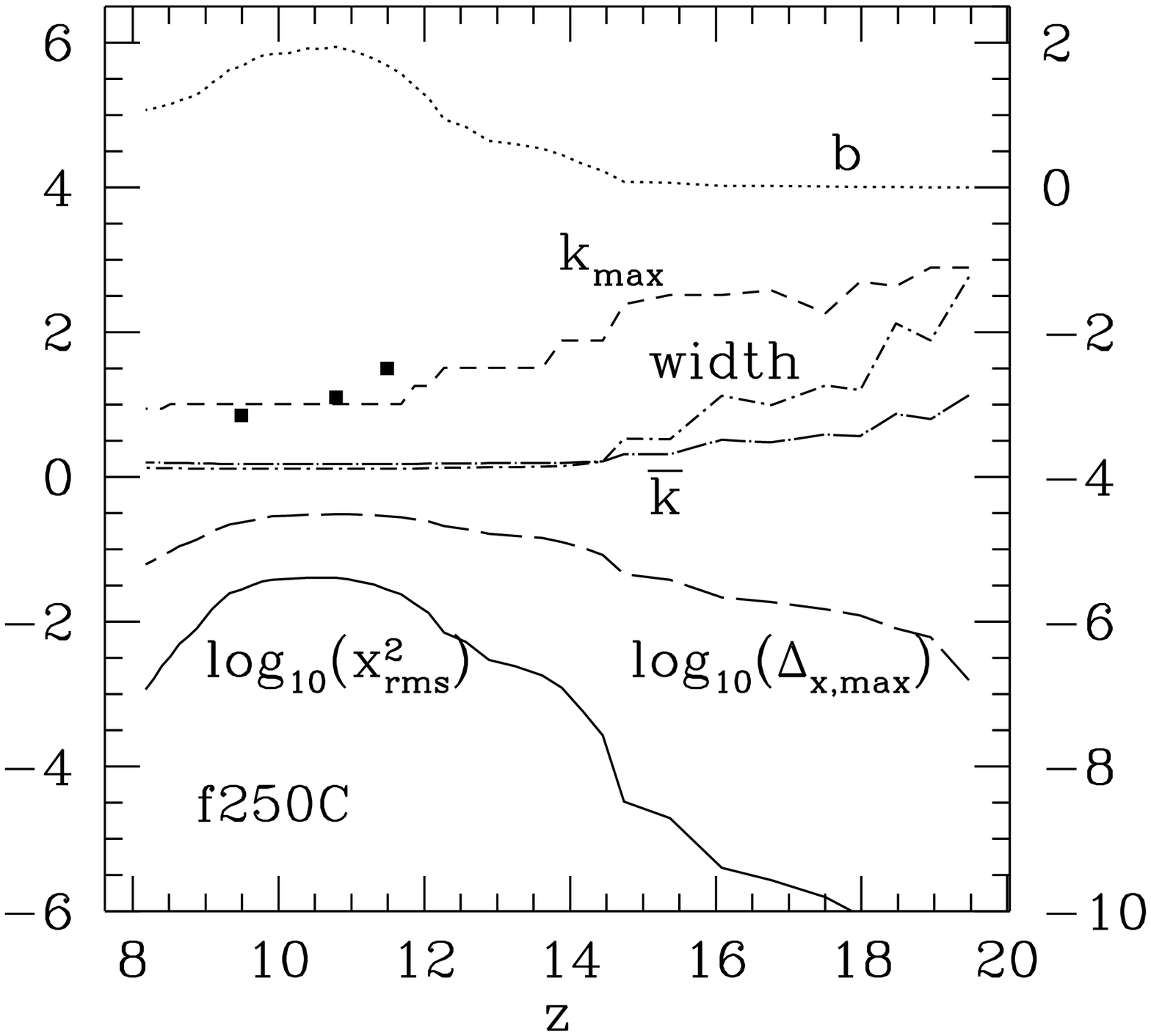}
\caption{\label{max_rms_fig} Evolution of $\Delta_x$. Plotted are
the peak value, $\Delta_{x,max}$, and the scale at which it occurs, 
$k_{\rm max}$, its effective width, and the corresponding rms, 
$x_{\rm rms}$ for all simulations, as labelled. We also show the bias, 
$b(z)$, at large scales of the ionized density with respect to the 
underlying density field (values indicated on the right axis) and 
$\bar{k}$ (see text for details). The values of $k_0/2$, where $k_0$ is 
the value of the characteristic ionized bubble scale which give a good 
fit to the $W(k)$ window function at large scales are also indicated 
for each case (black squares; see text for details).}
\end{figure*}

The characteristic scale corresponding to the peak of the power spectra in 
Figure~\ref{ps_w_ran_vel} is largely dictated by the typical scales of the
ionized bubbles, since, as we discussed in \S~\ref{lin_th_and_corr_sect}, the 
velocity and density are in the linear regime and do not have characteristic
scales within the range corresponding to our simulation volume. However, the 
ionized patches do have a typical scale. In Figure~\ref{power_spectra} we 
show $\Delta_x$, the power spectrum of the ionized fraction (which is the same 
as the one of the neutral fraction) for all of our simulation cases at the 
redshifts when the volume ionized fractions are $x_v=0.3, 0.5$ and $0.8$. In 
all cases the power spectra peak around wave-numbers $k\sim1\rm~h\,Mpc^{-1}$, 
which scale corresponds to $\ell\sim7000$. As reionization progresses, the 
peak moves gradually to slightly larger scales, resulting in a moderate 
widening of the peak of the kSZ temperature anisotropy sky power spectra 
and slightly less well-defined typical scale than seen in the single-redshift 
3D power spectra. In order to quantify the power spectra evolution better, in
Figure~\ref{max_rms_fig} we show the evolution vs. redshift of the peak 
amplitude, $\Delta_{x,max}$, the scale at which it occurs, $k_{\rm max}$, 
and the effective width of the peak, given by
\be
{\rm width}=\bar{k}\sinh\left\{\left[\overline{\delta(\ln k)^2}\right]^{1/2}\right\}
\ee
where
\be
\overline{\delta(\ln k)^2}=\frac{1}{x_{\rm rms}^2}
\int[(\ln k)^2-(\overline{\ln k})^2]\Delta_x^2d\ln k.
\ee
Here 
\be
\bar{k}=\exp(\overline{\ln k})
     =\exp\left[\frac{1}{x_{\rm rms}^2}\int(\ln k)\Delta_x^2d\ln k\right]
\ee
and 
\be 
x_{\rm rms}^2=\int \Delta^2_{x}d\ln(k),
\ee 
which gives the total integrated power per logarithmic interval. We also 
show $\bar{k}$ (in $h^{-1}$Mpc) and $x_{\rm rms}^2$. 
The amplitude of the peak initially rises, peaking at the point of maximum 
patchiness, $x_m\approx0.5$, and slightly decreases thereafter. The scale 
of the peak, $k_{\rm max}$,  (in $h^{-1}$Mpc) steadily increases in time, 
as the typical ionized regions grow. 
Approximate linear fits (in the form $a+bz$) to $k_{\rm max}(z)$ are given by 
$a=-0.528$~(-0.329,-1.135,-0.847) and $b=0.085$~(0.089,0.17,0.19) for 
simulation f2000 (f2000C, f250, f250C). The effective width of the power 
spectra and $\bar{k}$ are largely constant and roughly equal to each other, 
except for simulation f250C at early times. Finally, the rms initially 
strongly rises, peaking at slightly later times than the patchiness itself 
peaks (at $x_m\approx0.6$). 
 
\begin{figure*}
  \includegraphics[width=7in]{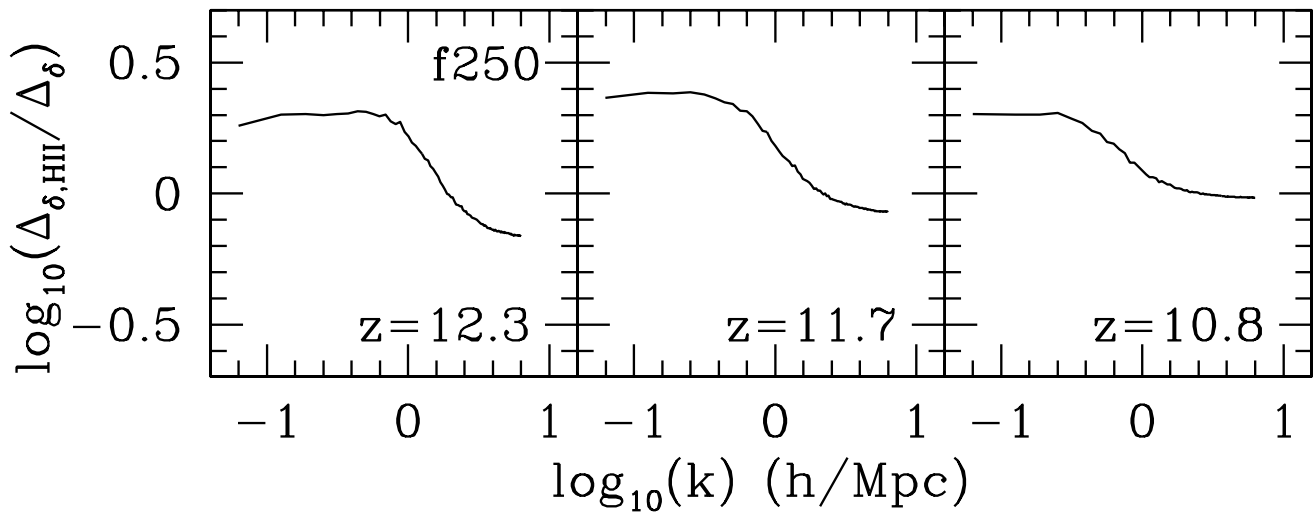}
\vspace{-10.5cm}
\caption{\label{power_spectra_ratios} Ratio of the power spectra of the 
ionized density and the total density, $\Delta_{\delta,HII}/\Delta_{\delta}$ 
for simulation f250 at volume ionized fractions $x_v=0.3$ (left), $x_v=0.5$ 
(middle), and $x_v=0.8$ (right). The corresponding redshifts are 
indicated on each panel.}
\end{figure*}
During the later stages of reionization ($x_m>0.3$), which contribute most 
of the kSZ signal the ionized density fluctuations at large scales are 
proportional to the density fluctuations with a bias coefficient $b(z)$ 
\citep[][see also Figure~\ref{power_spectra_ratios}. Note however that 
this is not the case at earlier times, when the ionized 
regions are relatively small and isolated and do not follow the large-scale
density perturbations.]{2006MNRAS.369.1625I}.
Based on this fact we can derive a handy approximate 
expression for the power spectrum of the ionized gas density, 
$\Delta_{\delta,HII}$, as follows. Let
\be
\Delta_{\delta,HII}(k,z)=b(z)W(k,z)\Delta_{\delta}(k,z),
\ee
where $W(k,z)$ is some window function asymptoting to 1 at large scales. 
We recognize that because of the zero or one nature of reionization (i.e. 
a region is either ionized or neutral when ionization fronts are 
unresolved, as is the case here), the random field $\delta n_e/\bar{n}_e$ is 
more complex than this relation between the power spectra indicates. 
However, it is reasonable expect that for low $k$, the electron density 
structure will have such a proportionality, with the window function being 
related to a quadratic superposition of form factors for the various H~II 
patches. If the reionization contribution from dense regions is lessened 
because of negative feedback from earlier ionizers on nearby galaxies, a 
situation we have ignored, the relation could be considerably modified. In 
spite of such complications, a simple relationship  characterized by a
characteristic scale related to the average size of the H~II regions
turns out to prevail as we now show.

The bias $b(z)$ is readily calculated from the simulation data (at the largest
scale available from our simulation volume) and is shown in 
Figure~\ref{max_rms_fig} (values shown on the right axis). The bias starts 
close to zero since very little of the gas is ionized, and has a peak value 
of 2-3, reached again around the time of maximum patchiness. For redshifts 
around the time of maximum patchiness (for mass-weighted ionized fraction 
$0.3<x_m<0.8$) and at large scales ($k\gtrsim1-2$) the window function 
$W(k,z)$ is reasonably well-fit (within factor of $\sim2$, and generally 
much better) by the following functional form:
\be
W(k,z)=\frac{1}{1+k^2/k_0^2(z)}
\label{window} 
\ee
where $k_0(z)$ is a characteristic scale parameter which gives the best fit
at the turnover. Samples of this window function, $b(z)W(k,z)$, for simulation
f250 are shown in Figure~\ref{power_spectra_ratios}. At small scales this 
function turns over and the above fit is not applicable. The values of 
$k_0(z)/2$ at $x_m=0.3,0.5$ and 0.8 (in $h^{-1}$Mpc) are shown in 
Figure~\ref{max_rms_fig} (squares). They are very close to the scale at 
which the ionized fraction power spectra peak, $k_{\rm max}$.

The addition of the missing large-scale velocity power outlined 
in the previous section boosts the sky power spectra at all scales, but 
especially at the largest ones, which further widens the peak. Nonetheless, 
the characteristic ionized bubble scale remains imprinted on the final 
results through their peak at $\ell\sim2000-9000$.

\begin{figure*}
  \includegraphics[width=3.5in]{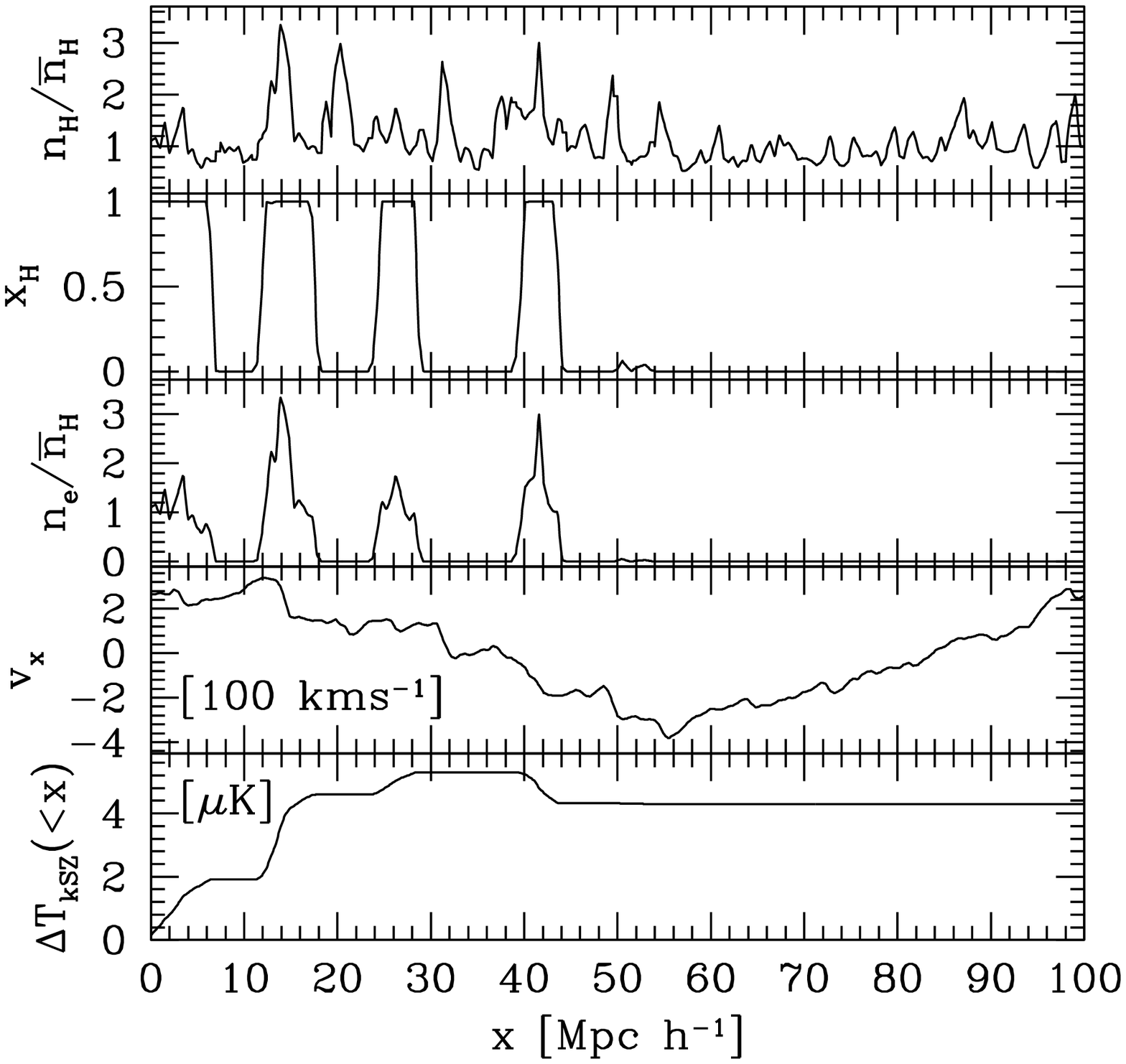}
  \includegraphics[width=3.5in]{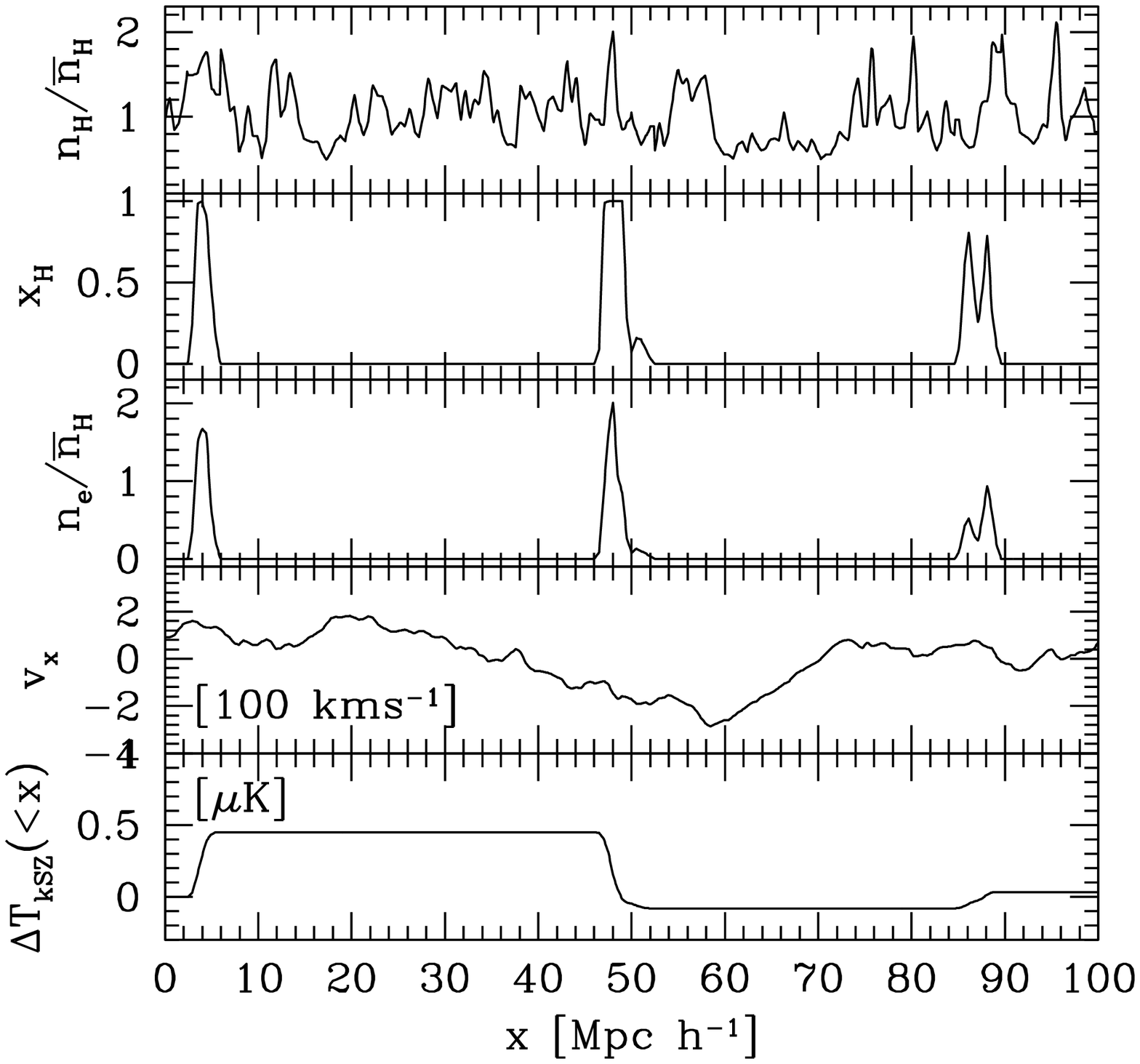}
\caption{\label{spectra_z13.62} LOS cuts through the simulation volume 
(simulation f250) at $z=13.622$ at which time the volume-weighted ionized 
fraction is $x_v=0.095$, and the mass-weighted one is $x_m=0.12$).}
\end{figure*}
\begin{figure}
  \includegraphics[width=3.5in]{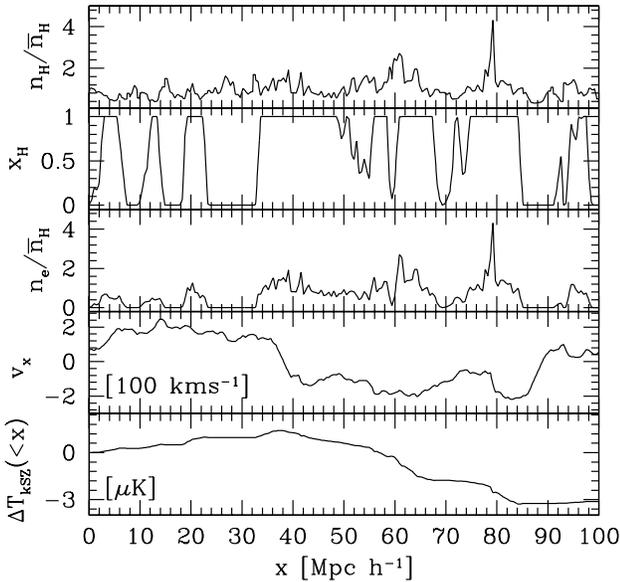}
\caption{\label{spectra_z11.31} LOS cuts through the simulation volume 
(simulation f250) at $z=11.310$  ($x_v=0.66$, $x_m=0.70$).}
\end{figure}

To gain some further insight in the complex nature of the derived kSZ signal, 
in Figures~\ref{spectra_z13.62} and \ref{spectra_z11.31} we show some sample
line-of-sight (LOS) cuts through the simulation volume (simulation f250) of the
density field in units of the mean, $n_H/\bar{n}_H$, ionized fraction, $x_H$,
electron number density in units of the mean gas density,
$n_e/\bar{n}_H=x_Hn_H/\bar{n}_H$, velocity component along that LOS, $v_x$,
and the accumulation of the kSZ temperature anisotropy integral (from left to
right), $\Delta T_{\rm kSZ}(<x)$. Early in the evolution ($z=13.62$;
Figure~\ref{spectra_z13.62}) the H~II regions tend to be associated with
high-density peaks, where the first halos (ionizing sources) form (inside-out
reionization scenario). The ionized fraction distribution acts as a filter, 
picking the density field portions that would contribute to the CMB
temperature anisotropies. The ionized regions are correlated with the
high-density peaks (inside-out reionization), but the correlation is not 
perfect, there are density peaks which still remain neutral, while at the same
time there are ionized regions which have a low overdensity, or are even 
underdense. The velocity field fluctuations are dominated by the largest
scales and are generally not closely correlated with the density field or the
ionized fraction. The velocity field determines the sign of the contribution
of a given ionized region to the total temperature anisotropy. A number of
different situations can be observed in our data. For example, in
Figure~\ref{spectra_z13.62} (left) the ionized regions are relatively closely
spaced together, separated by $\sim10$ Mpc, falling within the same
large-scale velocity fluctuation. All H~II regions except one are moving in
the positive direction, resulting in a relatively large integrated temperature 
anisotropy of $\sim4.4\,\mu K$. The other sample case from the same redshift 
(Figure~\ref{spectra_z13.62}, right) includes three ionized regions
along the LOS. These are separated by 40-50 comoving Mpc, and thus correspond 
to completely uncorrelated density fluctuations. The velocity field does not 
appear to correlate them, either, but nonetheless the temperature anisotropies 
they produce happen to almost exactly cancel each other, resulting in 
$\Delta T_{\rm kSZ}\approx 0$ along that LOS. At later times
(Figure~\ref{spectra_z11.31}) most of the volume is already ionized and each
LOS encounters a number of ionized regions, but ultimately almost all of the
total signal is contributed by the two highest-density peaks along the LOS,
both of which here happen to move in the negative direction, while the
contributions of the multiple other H~II regions largely cancel each
other. Once again, the density and velocity fields are not correlated. The
ionized fraction is less correlated with the high-density regions than at
early times. These examples demonstrate some of the complexities one has to
deal with when trying to derive the kSZ signal from patchy reionization
analytically, underscoring the need for large-scale simulations of this effect.

\section{Comparisons to previous work}

Previous simulations of the kSZ effect from patchy reionization 
\citep{2001ApJ...551....3G,2005MNRAS.360.1063S} predicted significantly lower
kSZ signals than the ones we find. Their power spectra reach maximum values
of $[\ell(\ell+1)C_\ell/(2\pi)]_{\rm max}\sim2\times10^{-14}$ and 
$\sim1.6\times10^{-13}$, respectively, compared to 
$[\ell(\ell+1)C_\ell/(2\pi)]_{\rm max}\sim1\times10^{-12}$ for our
simulations. This discrepancy is due to the small volumes used in these
simulations. As we showed in 
\S~\ref{lin_th_and_corr_sect} considering small volumes significantly reduces 
the power in the velocity field fluctuations. At the scale of the simulation 
box the bulk velocities are zero by definition, and the density is the mean 
one for the universe, thus any larger-scale fluctuations are not included, 
and the ones close to the box size are underestimated. The reionization
patchiness imprints its characteristic scale on the density and velocity
fluctuations, effectively smoothing the small-scale fluctuations below the
typical bubble sizes. On the other hand, the large-scale fluctuations still
should contribute to the kSZ signal, since the H~II regions are moving with
the large scale bulk motions. 

In addition to the missing large-scale power, these earlier reionization 
simulations also did not follow sufficient volumes to properly sample the 
size distribution of the ionized bubbles. The typical sizes of these ionized 
patches are of order 5-20 comoving Mpc, and become even larger at later times.
These large characteristic bubble scales are result of the strong clustering 
of ionizing sources at high redshifts. Capturing the full bubble size
distribution requires simulation volumes of order $(100\,\rm h^{-1}Mpc)^3$  
\citep{2006MNRAS.369.1625I}, and as a result the kSZ power spectra found by the 
smaller-box reionization simulations were largely flat, with no clear
characteristic scale. 

Several semi-analytical models for calculating the kSZ signature of
patchy reionization have been proposed in recent years
\citep{1998ApJ...508..435G,2003ApJ...598..756S,2005ApJ...630..643M}. 
\citet{1998ApJ...508..435G} proposed a very simple model, whereby the 
ionized patches are randomly distributed and have a given characteristic 
size $R$. Furthermore, they assumed that the density, velocity and ionization 
fraction fields are all uncorrelated with each other.  The ionized fraction
auto-correlation function is approximated as a Gaussian with rms given
by the (fixed) characteristic H~II region size. Under these
assumptions the kSZ power spectrum can be calculated analytically. The
quantitative details of the signal predicted by this model depend on
the assumptions made about the typical ionized patch size, and the
time and duration of reionization, but generically it predicts a
distribution which is very strongly peaked, more so than our
simulations find. This is related to the assumed Gaussian distribution
around a fixed characteristic size of the ionized patches. The actual
simulations also find a characteristic size for the ionized bubbles,
but one that is also evolving with redshift, and different size
distributions around it, which yields somewhat less sharp peaks than
this analytical model predicts.
  
\citet{2003ApJ...598..756S} proposed another simple model for the
reionization patchiness, which assumes that the correlation function
of the ionized patches is proportional to the one for the density
field, with a bias factor which is time-dependent, but not
scale-dependent. The power is filtered at small scales, below the
typical size of the ionized bubbles using a Gaussian filter in k-space.
The time-dependence of the typical size of the ionized patches is
assumed to be proportional to $[1-x_m(t)]^{-1/3}$, where $x_m(t)$ is
the mean mass-weighted ionized fraction at time $t$. The last function
is based on the semi-analytical models of reionization of
\citet{2003ApJ...595....1H}. The resulting kSZ power spectra are
fairly flat, with much less well-defined characteristic scale than our
simulation results, as should be expected based on the
quickly-evolving typical patch size assumed in this model. The peak
amplitudes of the results is somewhat higher than the ones we find, by
factors of $\sim2-3$, probably due to their assumed values of the
effective bias.

More recently, \citet{2005ApJ...630..643M} applied the 
semi-analytical reionization model of \citet{2004ApJ...613....1F} to
estimate the kSZ signal.  They found lower fluctuations in their
``single reionization episode'' models (by factor of $\sim3$),
although these models have similar mean reionization histories (i.e. 
evolution of the mean volume-weighted ionized fraction) to our
simulations. Their results also predict a peak at somewhat larger scales
($\ell\sim$2000) than ours ($\ell\sim$2000-9000). Their ``extended
reionization'' scenarios peak to similar values to the ones we find 
($[\ell(\ell+1)C_\ell/(2\pi)]_{\rm max}\sim10^{-12}$) again at 
$\ell\sim$2000, but have a very different assumed reionization histories 
than any of our simulations. They found that the kSZ post-reionization 
contribution dominates the patchy signal at the scales of interest, while we
find that the two signals are comparable in magnitude.

\section{Summary}

We have derived the kSZ CMB anisotropies due to the inhomogeneous reionization
of the universe. This is the first such calculation based on detailed,
large-scale radiative transfer simulations of this epoch. As we have shown above,
these simulations follow large enough volume to capture the full range of scales 
relevant to the large-scale reionization geometry. They also include most of the  
velocity fluctuations power, which is not the case for smaller-box simulations. 
We have also approximately corrected for the velocity power still missing from the 
box. The resulting sky power spectra peak at $\ell=2000-8000$ with maximum values 
of $[\ell(\ell+1)C_\ell/(2\pi)]_{\rm max}\sim1\times10^{-12}$. The angular scale 
of the peak roughly corresponds to the typical ionized bubble sizes observed 
in our simulations, which is $\sim5-20$~Mpc, depending on the assumed source 
efficiencies and the gas clumping at small scales. The kSZ anisotropy signal 
from reionization dominates the primary CMB signal above $\ell=3000$. At large 
scales the patchy kSZ signal depends largely on the source efficiencies and 
the large-scale velocity fields. It is
higher when sources are more efficient at producing ionizing photons, since
such sources produce large ionized regions, on average, than less efficient
sources. The introduction of sub-grid gas clumping in the radiative transfer 
simulations produce significantly more power at small scales, but has little
effect at large scales. The sub-grid gas clumping also significantly enhances 
the non-Gaussianity of the kSZ maps, resulting in up to one order of magnitude
more of the brightest kSZ regions than a Gaussian would predict. The
integrated kSZ signal is strong enough to be detected by upcoming experiments, 
like ACT and SPT. However, the separation of the patchy reionization signal
from the contribution by the fully-ionized gas after reionization seems
difficult. 

Our current simulations do not include the smallest atomically-cooling
ionizing sources, these with total mass $10^8\,M_\odot\lesssim M_{\rm
  tot}<2.5\times10^9\,M_\odot$, as these are not resolved in our base 
N-body simulations. These smaller sources have important effects on the early
global reionization history, but are also strongly suppressed due to
Jeans-mass filtering in the ionized regions \citep{selfregulated}. During 
most of the evolution the reionization process is thus dominated by the 
larger sources, which are resolved here, and which dictate the large-scale 
reionization geometry. Consequently, we do not expect that the presence of 
the low-mass
ionizing sources would change our current conclusions significantly.

\section*{Acknowledgments} 
This work was partially supported by NASA Astrophysical Theory Program grants
NAG5-10825 and NNG04G177G to PRS.

\bibliographystyle{apj}\bibliography{../refs}

\begin{thebibliography}{45}
\expandafter\ifx\csname natexlab\endcsname\relax\def\natexlab#1{#1}\fi

\bibitem[{{Box} \& {Muller}(1958)}]{Box-Muller}
{Box}, G.~E.~P. \& {Muller}, M.~E. 1958, Ann. Math. Stat., 29, 610

\bibitem[{{Bunker} {et~al.}(2006){Bunker}, {Stanway}, {Ellis}, {McMahon},
  {Eyles}, \& {Lacy}}]{2006NewAR..50...94B}
{Bunker}, A., {Stanway}, E., {Ellis}, R., {McMahon}, R., {Eyles}, L., \&
  {Lacy}, M. 2006, New Astronomy Review, 50, 94

\bibitem[{{Ciardi} \& {Madau}(2003)}]{2003ApJ...596....1C}
{Ciardi}, B. \& {Madau}, P. 2003, \apj, 596, 1

\bibitem[{{Ciardi} {et~al.}(2006){Ciardi}, {Scannapieco}, {Stoehr}, {Ferrara},
  {Iliev}, \& {Shapiro}}]{MH_sim}
{Ciardi}, B., {Scannapieco}, E., {Stoehr}, F., {Ferrara}, A., {Iliev}, I.~T.,
  \& {Shapiro}, P.~R. 2006, MNRAS, 366, 689

\bibitem[{{Furlanetto} {et~al.}(2006){Furlanetto}, {Oh}, \&
  {Briggs}}]{2006astro.ph..8032F}
{Furlanetto}, S., {Oh}, S.~P., \& {Briggs}, F. 2006, ArXiv Astrophysics
  e-prints (astro-ph/0608032)

\bibitem[{{Furlanetto} {et~al.}(2004{\natexlab{a}}){Furlanetto}, {Sokasian}, \&
  {Hernquist}}]{2004MNRAS.347..187F}
{Furlanetto}, S.~R., {Sokasian}, A., \& {Hernquist}, L. 2004{\natexlab{a}},
  \mnras, 347, 187

\bibitem[{{Furlanetto} {et~al.}(2004{\natexlab{b}}){Furlanetto}, {Zaldarriaga},
  \& {Hernquist}}]{2004ApJ...613....1F}
{Furlanetto}, S.~R., {Zaldarriaga}, M., \& {Hernquist}, L. 2004{\natexlab{b}},
  \apj, 613, 1

\bibitem[{{Gnedin} \& {Jaffe}(2001)}]{2001ApJ...551....3G}
{Gnedin}, N.~Y. \& {Jaffe}, A.~H. 2001, \apj, 551, 3

\bibitem[{{Gnedin} \& {Prada}(2004)}]{2004ApJ...608L..77G}
{Gnedin}, N.~Y. \& {Prada}, F. 2004, \apjl, 608, L77

\bibitem[{{Gruzinov} \& {Hu}(1998)}]{1998ApJ...508..435G}
{Gruzinov}, A. \& {Hu}, W. 1998, \apj, 508, 435

\bibitem[{{Haiman} \& {Holder}(2003)}]{2003ApJ...595....1H}
{Haiman}, Z. \& {Holder}, G.~P. 2003, \apj, 595, 1

\bibitem[{{Hu}(2000)}]{2000ApJ...529...12H}
{Hu}, W. 2000, \apj, 529, 12

\bibitem[{{Iliev} {et~al.}(2006{\natexlab{a}}){Iliev}, {Ciardi}, {Alvarez},
  {Maselli}, {Ferrara}, {Gnedin}, {Mellema}, {Nakamoto}, {Norman}, {Razoumov},
  {Rijkhorst}, {Ritzerveld}, {Shapiro}, {Susa}, {Umemura}, \&
  {Whalen}}]{2006MNRAS.tmp..873I}
{Iliev}, I.~T., {et~al.} 2006{\natexlab{a}}, \mnras, 371, 1057

\bibitem[{{Iliev} {et~al.}(2006{\natexlab{b}}){Iliev}, {Mellema}, {Pen},
  {Merz}, {Shapiro}, \& {Alvarez}}]{2006MNRAS.369.1625I}
{Iliev}, I.~T., {Mellema}, G., {Pen}, U.-L., {Merz}, H., {Shapiro}, P.~R., \&
  {Alvarez}, M.~A. 2006{\natexlab{b}}, \mnras, 369, 1625

\bibitem[{{Iliev} {et~al.}(2006{\natexlab{c}}){Iliev}, {Mellema}, {Shapiro}, \&
  {Pen}}]{selfregulated}
{Iliev}, I.~T., {Mellema}, G., {Shapiro}, P.~R., \& {Pen}, U.~L.
  2006{\natexlab{c}}, MNRAS, submitted (astro-ph/0607517)

\bibitem[{{Iliev} {et~al.}(2006{\natexlab{d}}){Iliev}, {Pen}, {Bond},
  {Mellema}, \& {Shapiro}}]{2006astro.ph..7209I}
{Iliev}, I.~T., {Pen}, U.-L., {Bond}, J.~R., {Mellema}, G., \& {Shapiro}, P.~R.
  2006{\natexlab{d}}, New Astronomy Reviews, in press (astro-ph/0607209)

\bibitem[{{Iliev} {et~al.}(2003){Iliev}, {Scannapieco}, {Martel}, \&
  {Shapiro}}]{2003MNRAS.341...81I}
{Iliev}, I.~T., {Scannapieco}, E., {Martel}, H., \& {Shapiro}, P.~R. 2003,
  \mnras, 341, 81

\bibitem[{{Iliev} {et~al.}(2005{\natexlab{a}}){Iliev}, {Scannapieco}, \&
  {Shapiro}}]{2005ApJ...624..491I}
{Iliev}, I.~T., {Scannapieco}, E., \& {Shapiro}, P.~R. 2005{\natexlab{a}},
  \apj, 624, 491

\bibitem[{{Iliev} {et~al.}(2002){Iliev}, {Shapiro}, {Ferrara}, \&
  {Martel}}]{2002ApJ...572L.123I}
{Iliev}, I.~T., {Shapiro}, P.~R., {Ferrara}, A., \& {Martel}, H. 2002, \apjl,
  572, L123

\bibitem[{{Iliev} {et~al.}(2005{\natexlab{b}}){Iliev}, {Shapiro}, \&
  {Raga}}]{2005MNRAS...361..405I}
{Iliev}, I.~T., {Shapiro}, P.~R., \& {Raga}, A.~C. 2005{\natexlab{b}}, \mnras,
  361, 405

\bibitem[{{Jaffe} \& {Kamionkowski}(1998)}]{1998PhRvD..58d3001J}
{Jaffe}, A.~H. \& {Kamionkowski}, M. 1998, \prd, 58, 043001

\bibitem[{{Kaiser}(1984)}]{1984ApJ...282..374K}
{Kaiser}, N. 1984, \apj, 282, 374

\bibitem[{{Ma} \& {Fry}(2002)}]{2002PhRvL..88u1301M}
{Ma}, C.-P. \& {Fry}, J.~N. 2002, Physical Review Letters, 88, 211301

\bibitem[{{Malhotra} \& {Rhoads}(2006)}]{2005astro.ph.11196M}
{Malhotra}, S. \& {Rhoads}, J. 2006, ApJ, 647, 95L

\bibitem[{{McQuinn} {et~al.}(2005){McQuinn}, {Furlanetto}, {Hernquist}, {Zahn},
  \& {Zaldarriaga}}]{2005ApJ...630..643M}
{McQuinn}, M., {Furlanetto}, S.~R., {Hernquist}, L., {Zahn}, O., \&
  {Zaldarriaga}, M. 2005, \apj, 630, 643

\bibitem[{{Mellema} {et~al.}(2006{\natexlab{a}}){Mellema}, {Iliev}, {Alvarez},
  \& {Shapiro}}]{methodpaper}
{Mellema}, G., {Iliev}, I.~T., {Alvarez}, M.~A., \& {Shapiro}, P.~R.
  2006{\natexlab{a}}, New Astronomy, 11, 374

\bibitem[{{Mellema} {et~al.}(2006{\natexlab{b}}){Mellema}, {Iliev}, {Pen}, \&
  {Shapiro}}]{21cmreionpaper}
{Mellema}, G., {Iliev}, I.~T., {Pen}, U.~L., \& {Shapiro}, P.~R.
  2006{\natexlab{b}}, MNRAS, in press (astro-ph/0603518)

\bibitem[{{Merz} {et~al.}(2005){Merz}, {Pen}, \& {Trac}}]{2005NewA...10..393M}
{Merz}, H., {Pen}, U.-L., \& {Trac}, H. 2005, New Astronomy, 10, 393

\bibitem[{{Ostriker} \& {Vishniac}(1986)}]{1986ApJ...306L..51O}
{Ostriker}, J.~P. \& {Vishniac}, E.~T. 1986, \apjl, 306, L51

\bibitem[{{Rhoads} {et~al.}(2003){Rhoads}, {Dey}, {Malhotra}, {Stern},
  {Spinrad}, {Jannuzi}, {Dawson}, {Brown}, \& {Landes}}]{2003AJ....125.1006R}
{Rhoads}, J.~E., {et~al.} 2003, \aj,
  125, 1006

\bibitem[{{Salvaterra} {et~al.}(2005){Salvaterra}, {Ciardi}, {Ferrara}, \&
  {Baccigalupi}}]{2005MNRAS.360.1063S}
{Salvaterra}, R., {Ciardi}, B., {Ferrara}, A., \& {Baccigalupi}, C. 2005,
  \mnras, 360, 1063

\bibitem[{{Santos} {et~al.}(2003){Santos}, {Cooray}, {Haiman}, {Knox}, \&
  {Ma}}]{2003ApJ...598..756S}
{Santos}, M.~G., {Cooray}, A., {Haiman}, Z., {Knox}, L., \& {Ma}, C.-P. 2003,
  \apj, 598, 756

\bibitem[{{Scott} \& {Rees}(1990)}]{1990MNRAS.247..510S}
{Scott}, D. \& {Rees}, M.~J. 1990, \mnras, 247, 510

\bibitem[{{Seljak} \& {Zaldarriaga}(1996)}]{1996ApJ...469..437S}
{Seljak}, U. \& {Zaldarriaga}, M. 1996, \apj, 469, 437

\bibitem[{{Shapiro} {et~al.}(2006){Shapiro}, {Ahn}, {Alvarez}, {Iliev},
  {Martel}, \& {Ryu}}]{2006ApJ...646..681S}
{Shapiro}, P.~R., {Ahn}, K., {Alvarez}, M.~A., {Iliev}, I.~T., {Martel}, H., \&
  {Ryu}, D. 2006, \apj, 646, 681

\bibitem[{{Shapiro} {et~al.}(2004){Shapiro}, {Iliev}, \&
  {Raga}}]{2004MNRAS.348..753S}
{Shapiro}, P.~R., {Iliev}, I.~T., \& {Raga}, A.~C. 2004, \mnras, 348, 753

\bibitem[{{Spergel} {et~al.}(2003)}]{2003ApJS..148..175S}
{Spergel}, D.~N. et~al. 2003, \apjs, 148, 175

\bibitem[{{Springel} {et~al.}(2001){Springel}, {White}, \&
  {Hernquist}}]{2001ApJ...549..681S}
{Springel}, V., {White}, M., \& {Hernquist}, L. 2001, \apj, 549, 681

\bibitem[{{Stanway} \& {et al.}(2004)}]{2004ApJ...604L..13S}
{Stanway}, E.~R. \& {et al.} 2004, \apjl, 604, L13

\bibitem[{{Sunyaev} \& {Zeldovich}(1980)}]{1980MNRAS.190..413S}
{Sunyaev}, R.~A. \& {Zeldovich}, I.~B. 1980, \mnras, 190, 413

\bibitem[{{Tozzi} {et~al.}(2000){Tozzi}, {Madau}, {Meiksin}, \&
  {Rees}}]{2000ApJ...528..597T}
{Tozzi}, P., {Madau}, P., {Meiksin}, A., \& {Rees}, M.~J. 2000, \apj, 528, 597

\bibitem[{{Vishniac}(1987)}]{1987ApJ...322..597V}
{Vishniac}, E.~T. 1987, \apj, 322, 597

\bibitem[{{Zahn} {et~al.}(2005){Zahn}, {Zaldarriaga}, {Hernquist}, \&
  {McQuinn}}]{2005ApJ...630..657Z}
{Zahn}, O., {Zaldarriaga}, M., {Hernquist}, L., \& {McQuinn}, M. 2005, \apj,
  630, 657

\bibitem[{{Zeldovich} \& {Sunyaev}(1969)}]{1969Ap&SS...4..301Z}
{Zeldovich}, Y.~B. \& {Sunyaev}, R.~A. 1969, \apss, 4, 301

\bibitem[{{Zhang} {et~al.}(2004){Zhang}, {Pen}, \&
  {Trac}}]{2004MNRAS.347.1224Z}
{Zhang}, P., {Pen}, U.-L., \& {Trac}, H. 2004, \mnras, 347, 1224

\end{thebibliography}

\end{document}